\documentclass[12pt]{article}

\usepackage{packages_in}

\usepackage{eqtabular}

\usepackage{colornames}

\usepackage{terminology_in}

\usepackage{style_in}

\begin{document}


\maketitle

\renewcommand{\thepage}{\arabic{page}}
\setcounter{page}{1}

\thispagestyle{empty}
\begin{center} \today \end{center}

\begin{abstract} 
\noindent This article considers a stable vector autoregressive (VAR) model and investigates return predictability in a Bayesian context. The VAR system comprises asset returns and the dividend-price ratio as proposed in \citet{Cochrane_2008}, and allows pinning down the question of return predictability to the value of one particular model parameter. We develop a new shrinkage type prior for this parameter and compare our Bayesian approach to ordinary least squares estimation and to the reduced-bias estimator proposed in \citet{Amihud_2004}. A simulation study shows that the Bayesian approach dominates the reduced-bias estimator in terms of observed size (false positive) and power (false negative). We apply our methodology to annual CRSP value-weighted returns running, respectively, from 1926 to 2004 and from 1953 to 2021. For the first sample, the Bayesian approach supports the hypothesis of no return predictability, while for the second data set weak evidence for predictability is observed.
\end{abstract}

\begin{codes}  \end{codes}


\clearpage

\section{Introduction} \label{sec:intro}
The question whether stock market returns are predictable has been extensively studied in the empirical financial literature. Famously \citet{Fama_1970} postulated that financial markets are informationally efficient, meaning that all available information is already reflected in stock market prices. This implies that in any predictive regression for stock market returns, we expect an insignificant and close to zero regression coefficient for any regressor. Nevertheless, empirical literature \citep[see, e.g.,][]{Fama_French_1988, Campbell_1987} indicates that some financial variables (e.g., the dividend-price ratio) predict stock market returns, contradicting the efficient market hypothesis. Asset return predictability has been discussed in various strands of financial literature \citep[for an overview see e.g.][Chapter~5]{Campbell_book_2017} as well as in the machine learning and statistical literature \citep[for an overview see e.g.][]{nagel2021machine}. A prominent example is \citet{Welch_2008}, where the authors demonstrate that past  average stock returns are superior in predicting the current return compared to explanatory variables constructed by the authors \citep[for an overview see e.g.][]{Golez_2018}.

The main focus of this article is on parameter estimation and testing in linear models and their relation to asset return predictability. In particular, our work is related to the seminal paper by 
\citet{Stambaugh_1999}, who pointed out that some of the proposed regressors (e.g., the dividend-price ratio) are persistent, which leads -- when applying least squares estimation -- to incorrect inference and biased estimates. Several ways were already proposed to address this problem. For example, \citet{Stambaugh_1999} employed a Bayesian approach and found mixed evidence for the predictability. The results depend on the time frame and a combination of a likelihood and a prior.

\citet{Cochrane_2008} applies the \citet{Campbell_1988} log-linearization for stock market returns, and obtains a joint VAR for the log asset returns, the log dividend-price ratio and log dividend growth. By estimating the model parameters by ordinary least squares, \citet{Cochrane_2008} concludes that we cannot reject the null hypothesis of no predictability at the 5\% level. However, by constructing the joint null hypothesis for returns and dividend growth, the absence of dividend growth predictability (by the dividend-price ratio) implies that the dividend-price ratio predicts future stock market returns. In this article, we follow \citet{Cochrane_2008} and consider a bivariate autoregressive (VAR) system comprising asset returns and the lagged dividend-price ratio. This simple model allows pinning down the question of return predictability to the value of one particular parameter for the VAR, denoted $\beta$ in the following.

As already pointed out in \citet{Kendall1954}, since the predication variable is predetermined the ordinary least squares estimator of the autoregressive parameter of an AR(1) model, denoted $\phi$ in this article, is biased. 
In addition, as shown in 
\citet{Stambaugh_1999} or 
\citet{Phillips_2015}[and literature cited therein],
due to this bias in the estimate of this autoregressive parameter, the estimate of the parameter $\beta$ 
in a regression of returns on the lagged dividend-price ratio becomes biased. 

To cope with these problems and to reduce both biases, this article develops a Bayesian estimation approach that is strongly influenced by the many pioneering contributions of Herman van Dijk in the area of Bayesian econometrics. As shown by Herman van Dijk and coauthors \citep[see, e.g.,][]{dep-etal:bay,bas-etal:bay},
 Bayesian inference for systems of equations can be very challenging 
and requires a careful choice of an appropriate representation of suitable prior distributions and of efficient techniques to sample from the corresponding posterior distribution. 
In the present article, we use the Bayesian control function approach  \citep[see, e.g.,][]{Lopes_2014},
as a representation of the bivariate predictive system of \citet{Cochrane_2008} and develop a new prior distribution and a new posterior sampling technique to perform Bayesian inference.

In a time series context, the prior on the autoregressive coefficient $\phi$ can be particularly influential, as shown in the seminal work by Herman van Dijk \citep{Schotman_1991,sch-van:bay_ana}. 
In the present paper, a prior based on 
\citet{Berger_1994} is applied to improve the parameter estimates for the autoregressive parameter $\phi$. To reduce the bias for the estimator of $\beta$ we propose a shrinkage prior for $\beta$ based on a prior on the  (population) coefficient of determination $R^2$. As observed in \citet{Cadonna_2020,Giannone_2021,Zhang_2020_r2_d2} imposing a Beta prior on $R^2$ results in a shrinkage prior for the parameter $\beta$ known as triple gamma prior or normal-gamma-gamma prior. In addition, we calculate the Bayes factor to test the null hypothesis $\beta=0$ against the alternative $\beta \not=0$. To improve the performance of the Bayesian test, measured in terms of the {estimated false negative} (observed power) and {estimated false positive rates} (observed size), we additionally impose a hyperprior on the first parameter of the Beta distribution on $R^2$. 

Together with coauthors, Herman van Dijk made outstanding contributions regarding computational techniques for systems of equations and developed various Monte Carlo techniques for efficient sampling from the posterior distribution, \citep[see, e.g.,][]{klo-van:bay,hoo-etal:sha,zel-etal:bay_ana}. In the present paper, we derive a new
Markov Chain Monte Carlo (MCMC) sampler to sample efficiently from the posterior distribution of the bivariate predictive system of \citet{Cochrane_2008} under our specific prior choices.

In a simulation study, we compare the performance of our Bayesian approach to the ordinary least squares estimator (OLS) and the reduced-bias estimator  (RBE) proposed in \citet{Amihud_2004}.  
In terms of root mean square error (RMSE) and mean absolute error (MAE),  the Bayesian approach proposed in this article leads to superior performance relative to ordinary least squares estimation, and in most settings also compared to the reduced-bias approach. Regarding testing, due to the bias observed with OLS estimation, we get strong oversizing but high power with OLS. Oversizing is slightly reduced with the reduced-bias estimator, but the observed power is quite low. For our Bayesian approach, the prior parameters are chosen such that the false positive rate 
becomes quite close to the (frequentistic) theoretical size. With this setting, we get a false negative rate which is only slightly larger than with OLS and clearly outperforms the reduced-bias estimator.

In our empirical part, we analyse two data sets with annual CRSP value-weighted returns and corresponding dividend-price ratios. The first data set contains annual data from January 1926 until December 2004. We use this period to have comparable data with \citet{Cochrane_2008}. The second data set contains the post-war data from January 1953 until December 2021. We use this period to analyse 
also more recent data.  OLS suggests that for both data sets we have predictability on a level larger than $6.2\%$, while the RBE suggests that we have no predictability at usual significance levels ($p$-values larger than 19\%). The Bayesian tests suggest that we have no predictability for the first sample, however, we have weak evidence for predictability in the second sample.

This paper is organized as follows: Section~\ref{sec:model}
presents the model. 
Section~\ref{sec:bayesian_analysis} discusses Bayesian estimation and provides details 
on prior choices and posterior inference using Markov Chain Monte Carlo (MCMC) methods.
Section~\ref{sec:simulation} provides the simulation results, while Section~\ref{sec:financial} applies the Bayesian sampler to empirical financial data. Finally, Section~\ref{sec:conclusion} concludes.

\section{Model and Asset Predictability} \label{sec:model}

Let $D_t$ and $P_t$ abbreviate the dividend payment and the asset price in period $t$, $y_t := (P_t + D_t)/P_{t-1}$ denote  the asset return from period $t-1$ to $t$ and $x_t := \log D_t  - \log P_t$ the log dividend-price ratio.
This article follows \citet{Cochrane_2008} and considers the following restricted  vector autoregressive (VAR) system:%
{\footnote{In this article, we apply the following notation:
$\Betadis{.,.}$
$\Betapr{.,.} $,
$\Gammad{.,.}$
$\Gammainv{.,.}$
$\mathcal{IW}_p(.,.)$
$\Normal{.,.}$,
$\Fd{.,.}$ denote the Beta, the Beta-prime, the Gamma, the inverse Gamma, the inverse Wishart, the Normal and the F-distribution, respectively. 
For matrices and vectors, we use bold-face notation. If not otherwise stated, the vectors considered are column vectors,
$\mathbf{0}_p$, $\mathbf{1}_p$ denote the p-dimensional vector of zeros and one, $\mathbf{I}_p$ is the p-dimensional identify matrix and $\mathds{1}_{\{ \cdot \} }$ abbreviates an indicator function. Finally, $\log(z)$ denotes the logarithmus naturalis, that is the logarithm with basis $e$. } }
\begin{align}
	\left( 
	\begin{array}{c}
		x_{t} \\
		y_{t}
		\end{array} \right) &=
	\left( 
	\begin{array}{c}
		\alpX \\
		\alpY
	\end{array} \right) +
	\left( 
\begin{array}{cc}
	\phi & 0 \\
	\beta & 0
\end{array} \right) 
	\left( 
\begin{array}{c}
	x_{t-1} \\
	y_{t-1}
\end{array} \right) 
+
	\left( 
\begin{array}{c}
	\epsilon_t^x \\
	\epsilon_t^y
\end{array} \right) \ .
\label{eq:model_1}
\end{align}
This restricted VAR model contains a first-order auto-regressive process (including an intercept term) for the log dividend-price ratio, and a regression of asset returns on the lagged log dividend-price ratio. We call the first and the second equation of model~(\ref{eq:model_1}), {\em dividend-price ratio equation} and {\em return equation}, respectively. A non-zero regression parameter $\beta$ implies return predictability. 
We assume that $|\phi|<1$, resulting in a stable system;
see Appendix~\ref{app:ar_1.phi} for a motivation for this choice. 
The errors $(\epsilon_t^x, \epsilon_t^y)^{\prime}$ are serially independent and identically distributed as bivariate normal random variables with mean zero and full rank covariance matrix $\mathbf{\Sigma}$:
\begin{align}
\left( \begin{array}{c}
	\epsilon_t^x \\
	\epsilon_t^y
\end{array} \right)
\sim  \mathcal{N}\left( \mathbf{0}_{2} , \mathbf{\Sigma}  \right),
\qquad
\mathbf{\Sigma} := \left( 
\begin{array}{cc}
\sigma_{x}^2 & \sigma_{xy} \\
\sigma_{xy} & \sigma_{y}^2
\end{array}	
\right). \label{eq:model_1A}
\end{align}
By this assumption, the instantaneous correlation between the innovations of the dividend-price ratio and the asset returns is allowed. Additionally assuming that $\Delta \log D_{t}:= \log D_t - \log D_{t-1} = \alpha^d + \beta^d x_{t-1}  + \epsilon^d_{t}$, using the \citet{Campbell_1988} linearization, \citet{Cochrane_2008} {has shown} that $\beta \approx 1 + \beta^d - \rho \phi$, where $ 0< \rho <1$ is the average dividend-price ratio about which one linearizes. Since $|\rho \phi|<1$, $\beta^d = 0$ implies $\beta \not=0$ and vice versa. Hence, if dividends are not predictable, then returns have to be predictable (and vice versa). 

\section{Bayesian Analysis} \label{sec:bayesian_analysis}

Before we develop Bayesian tools for parameter estimation, let us remark on ordinary least squares  estimation applied to the system (\ref{eq:model_1}).
Let $\widehat{\phi}^{OLS}$ and $\widehat{\beta}^{OLS}$ denote the OLS estimators of, respectively,  $\phi$ and $\beta$ obtained from the observed data $\{ (x_t)_{t=0,\dots,T}, (y_t)_{t=1,\dots,T} \}$. In his seminal paper, \citet{Stambaugh_1999} shows that OLS estimation produces the following bias:
\begin{equation} \label{eq:bias}
	\mathbb{E}\left( \widehat{\beta}^{OLS} - \beta \right) = \frac{\sigma_{xy}}{\sigma_{x}^2} \mathbb{E} \left( \widehat{\phi}^{OLS} - \phi \right).
\end{equation}
The serial correlation of the dividend-price ratio is high in empirical data sets \citep[e.g.,][obtains $\widehat{\phi}^{OLS}$ equal to 0.941]{Cochrane_2008}. As already demonstrated by \citet{Kendall1954}, 
\begin{align*}
    \mathbb{E} \left( \widehat{\phi}^{OLS} - \phi \right) \approx - \frac{1 + 3 \phi}{T} < 0,
\end{align*}
where $T$ is the number of observations. Combined with $ \sigma_{xy} < 0$ (e.g., \citet{Cochrane_2008} obtained $\widehat{\sigma}_{xy} = -0.7$) upward biased estimates of $\beta$ can be expected when applying OLS. 

There are several ideas to address this problem \citep[for an overview see e.g.][p.~144-148]{Campbell_book_2017}. \citet{Amihud_2004} developed a reduced-bias estimator\footnote{The authors propose two reduced-bias estimators for the autoregressive coefficient $\phi$  denoted, respectively, by $\widehat{\phi}^{RBE}$ and $\widehat{\phi}^{RBE,1}$, 
where $\widehat{\phi}^{RBE}$ is provided in (\ref{eq:RBE}) and $\widehat{\phi}^{RBE,1}$ does not include the last term in (\ref{eq:RBE}).} that is used to reduce the bias of the OLS estimate $\widehat{\phi}^{OLS}$ and is
defined by:
\begin{align}
	\widehat{\phi}^{RBE} &= \widehat{\phi}^{OLS} + \frac{1 + 3\widehat{\phi}^{OLS}}{T} + 3\frac{1 + 3\widehat{\phi}^{OLS}}{T^2} \label{eq:RBE}.
\end{align}
Then, by using the parameter estimates for the dividend-price ratio equation of model~(\ref{eq:model_1}), the residuals $\widehat{\epsilon_t}^{x,RBE}$, $t=1,\dots,T$, are obtained. Finally, the residuals $\widehat{\epsilon_t}^{x,RBE}$ are used as an additional regressor in the return equation in system~(\ref{eq:model_1}). \citet{Amihud_2004}[Lemma~1,  Theorem~4, and Section~B] {provide infeasible estimators of $\phi$ and $\beta$ which are unbiased and show that the corresponding feasible estimators are reduced-bias}. 

In this article, the Bayesian control function approach  \citep[see, e.g.,][]{Lopes_2014},
is applied to system~(\ref{eq:model_1}) and adjusted to a time series context to account for the biases discussed above. By the properties of the normal distribution \citep[see, e.g.,][Lemma~10.4]{Ruudbook2000} we get
\begin{align*}
    \epsilon_t^y | 
\epsilon_t^x 
\sim \mathcal{N}
\left( 
\frac{\sigma_{xy}}{\sigma_{x}^2}
\epsilon_t^x , 
\sigma_{y}^2 - \frac{\sigma_{xy}^2}{\sigma_{x}^2}
\right).
\end{align*}
This allows to describe $(x_t,y_t)^{\prime}$ by means of the following representation:
\begin{align}\label{eq:model_3}
	\left( 
	\begin{array}{c}
		x_{t} \\
		y_{t}
		\end{array} \right) &=
	\left( 
	\begin{array}{c}
		\alpX \\
		\alpY
	\end{array} \right) +
	\left( 
\begin{array}{ccc}
	\phi & 0 & 0\\
	\beta & 0 & \psi
\end{array} \right) 
	\left( 
\begin{array}{c}
	x_{t-1} \\
	y_{t-1} \\
	\epsilon_t^x
\end{array} \right) 
+
	\left( 
\begin{array}{c}
	\epsilon_t^x \\
	\tilde \epsilon_t^y
\end{array} \right) \ , \nonumber \\
& 
\left(
\begin{array}{c}
	\epsilon_t^x \\
	\tilde \epsilon_t^y
\end{array} \right)
\sim  \mathcal{N}\left( \mathbf{0}_{2} , 
\left( \begin{array}{cc}
	\sigma_{x}^2 & 0 \\ 
	0 & \sigmacon
\end{array} \right) \right), 
\end{align}
with a one-to-one mapping to system (\ref{eq:model_1}) given by $\sigma_{xy}=\psi \sigma_{x}^2$ and $\sigma_{y}^2=\sigmacon + \sigma_x^2\psi^2$.
In  representation (\ref{eq:model_3}), the innovations $\epsilon_t^x$ from the dividend-price ratio equation are used as an additional regressor in the return equation with regression coefficient $\psi$, making the innovations $\tilde \epsilon_t^y$ being independent of $\epsilon_t^x$.
While both representations would yield identical estimators in a maximum likelihood framework, changing parameterization in Bayesian inference allows choosing different priors. 

For the  observed data we apply the following notation: 
$\mathbf{x} := (x_t)_{t=1,\dots,T}$, $\mathbf{y} := (y_t)_{t=1,\dots,T}$ and  $\mathbb{D}_T := \{ x_0,
\mathbf{x}, \mathbf{y}\}$. We assume that the initial dividend-price ratio $x_0$ is observed, however, we do not condition on $x_0$ in our analysis. Following \citet{poi:eff} and \citet{Stambaugh_1999}, we work with the unconditional likelihood and assume that $x_0$ is a realization from the stationary distribution of the AR(1) process  defined in the dividend-price ratio equation. Hence, $p(x_0| \alpX, \phi, \sigma_{x}^2)$ is a normal density with mean $\alpX/(1-\phi)$
and variance $\sigma_{x}^2/(1 - \phi^2)$. 

With $\bm{\theta} 
:= \{ \alpX, \alpY, \phi, \beta, \sigma_{x}^2,
\psi, \sigmacon \}
$ collecting all model parameters in (\ref{eq:model_3}), 
we combine  the conditional likelihood $p(\mathbf{y}, \mathbf{x} |x_0, {\bm \theta} )$ with the initial distribution $ p(x_0| \alpX, \phi, \sigma_{x}^2)$, to define the exact likelihood
$p \left( \mathbf{x}, \mathbf{y}, x_0| {\bm \theta} \right)  
	= p \left(\mathbf{x} , \mathbf{y}|x_0, {\bm \theta} \right) p \left( x_0| \alpX, \phi, \sigma_{x}^2 \right) $:
\begin{eqnarray}
  &  p \left( \mathbf{x}, \mathbf{y}, x_0| {\bm \theta} \right)  
	= p \left( \mathbf{y}|\mathbf{x}, \alpY, \beta, \sigmacon, \psi  \right) 
	p \left(\mathbf{x} |x_0, \alpX, \phi, \sigma_{x}^2 \right) 
	p \left( x_0| \alpX, \phi, \sigma_{x}^2 \right)
		&
  \label{eq:exact_likelihood} \\
	& \displaystyle = \prod_{t=1}^{T} \frac{1}{ {\sqrt{
	 2\pi \sigmacon   }}}
	 \exp 
	 \left( -\frac{1}{ 2 
	 \sigmacon }
	 \left( y_{t} - \alpY - \beta x_{t-1} -
	 \psi \epsilon_t^x \right)^2  \right) &
	 \nonumber 
  \\ & 
	 \displaystyle \times 
	 \prod_{t=1}^{T} \frac{1}{ {\sqrt{
	 {2\pi \sigma_{x}^2  } } } }
	 \exp 
	 \left( -\frac{1}{2 
	 \sigma_{x}^2 }
	 \left( x_{t} - \alpX - \phi x_{t-1} 
	  \right)^2 \right)
	 \nonumber 
	 \times \sqrt{\frac{1 - \phi^2}{2\pi\sigma_{x}^2}}  \exp\left( -\frac{1 - \phi^ 2}{2\sigma_{x}^2} \left( x_0 - 
  \frac{\alpha^{x} }{1- \phi}
  \right)^2 \right) \; . & \nonumber
\end{eqnarray}
This exact likelihood is combined with a prior $p(\bm{\theta})$ to define the posterior distribution  $p(\bm{\theta}|\mathbb{D}_T)$.
Our specific prior choice is discussed in Section~\ref{sec:priors}, while sampling from the posterior distribution $p(\bm{\theta}|\mathbb{D}_T)$ is discussed in  Section~\ref{sec:mcmc}.

\subsection{Priors} \label{sec:priors}

The prior distribution $p(\bm{\theta}) $ of $\bm{\theta} 
= \{ \alpX, \alpY, \phi, \beta, \sigma_{x}^2,
\psi, \sigmacon \}
$ is assumed to take the form
\begin{align}
p(\bm{\theta}) 
:= p(\alpX| \phi) p (\alpY) p(\phi) p(\beta| \phi,\sigma_{x}^2, \psi, \sigmacon) p(\sigma_{x}^2)  p(\psi) p(\sigmacon) \;  .
\end{align}
Regarding the parameters $\sigma_{x}^2$ , $\psi$ and $\sigmacon$ defining the covariance matrix $\bm{\Sigma}$, 
we apply the Cholesky-based prior proposed by \citet{Lopes_2014} 
by choosing independent inverse gamma prior distributions for $\sigma_{x}^2$ and $\sigmacon$ and a normal prior  for $\psi$: 
\begin{align}
	\sigma_{x}^2  &\sim \mathcal{IG}(\nuX, 
 \SX)
	\ , \ \
	\sigmacon  \sim \mathcal{IG}(\nuY, \SY) 
	\ ,
 \ \
	\psi \sim \mathcal{N}(\mu_0^{\psi}, \Sigma_0^{\psi})\label{eq:prior_variance_covariance_cholesky} \ .
\end{align} 
As shown by \citet{Lopes_2014}, the Cholesky-based prior implies a much more flexible prior on $\bm{\Sigma}$ than the commonly applied inverse Wishart prior $\bm{\Sigma} \sim \mathcal{IW}_2(\nu_0, \bm{S}_0)$,
see e.g. \citet{Rossi_book_2006}. The inverse Wishart prior exhibits less flexible behaviour because of a single tightness parameter $\nu_0$ and implies an implicit prior on $\psi$ that might have unintended effects on the posterior distribution of the parameters of interest.

We deviate from the prior choices in \citet{Lopes_2014} regarding (most of) the other model parameters for two reasons. First, we are dealing with an AR(1) process in the first stage equation and this impacts our choice of priors on $\phi$ and $\alpX$. There is a vast literature on priors for autoregressive processes \citep[see, e.g.,][]{Schotman_1991, Kastner_2014,Choi_book_2015, Krone_2017} and the main conclusion is that the prior choice matters a lot. 
As discussed in Section~\ref{sec:model}, we assume that { $\{x_t\}_{t \in \mathbb{Z} }$ is a stationary process which requires
$\phi \in (-1,1)$}. Following 
the arguments in \citet{Stambaugh_1999}, we demand that $\phi \in [0,1)$  and apply  the reference prior of  \citet{Berger_1994}, truncated to the support  $[0,1)$, as \citet{Stambaugh_1999} do. This yields the following prior distribution for $\phi$:
\begin{equation}
	p(\phi) = \frac{2}{\pi\sqrt{1 - \phi^2}}
 \mathds{1}_{\{ \phi \in [0,1) \} } \ . \label{eq:prior_reference}
\end{equation}
Assuming that the stationary mean  $\muX \sim \mathcal{N}\left(\mualpX, \SalpX\right)$ of $\left( x_t \right)_{t \in \mathbb{Z}} $ apriori is independent of $\phi$, we obtain the following prior for  $\alpX$ conditional on $\phi$:
\begin{equation}\label{eq:prior_alpha_x}
	\alpX|\phi \sim \mathcal{N}\left(\mualpX(1 - \phi), \SalpX(1 - \phi)^2\right).
\end{equation}
Regarding the regression coefficient $\beta$ and the intercept $\alpY$ the return equation, the standard choice is an independence prior, i.e.:
\begin{align}\label{eq:prior_beta_alpha}
    \beta &\sim \mathcal{N}(\mu_0^{\beta}, \Sigma_0^{\beta}) \ , \ \
    \text{ and } \ \
    \alpha^{y} \sim \mathcal{N}(\mu_0^{\alpY}, \Sigma_0^{\alpY}) \ .
\end{align}
Following earlier work by \citet{poi:pri}, \citet{wac-war:pre} and \citet{Wachter_2015}, we construct an implicit prior for $\beta$ based on choosing a proper prior on the coefficient of determination $R^2$ in the return equation. Similarly to \citet{Giannone_2021} and \citet{Zhang_2020_r2_d2}, we impose a Beta-prior on {the population} $R^2$, that is:
\begin{align} \label{eq:prior_R2}
    R^2 \sim \mathcal{B} \left( \aR, \bR \right).
\end{align}
Such a design allows incorporating prior knowledge that the predictability of returns is, in general weak, and allows to elicit informed prior choices based on an investigator's perception of how large $\Rsquare$ is going to be.

By using the relationship between the Beta distribution and the Beta prime distribution, we obtain a shrinkage prior for the parameter $\beta$ known as triple gamma prior or normal-gamma-gamma prior \citep[see, e.g.,][]{Johnson_book_1995,Cadonna_2020}.  Appendix~\ref{app:explicit_prior_r2} shows that in this case the prior on $\beta$ has the following hierarchical representation:
\begin{eqnarray} \label{eq:prior_R2D2}
	\beta |  \phi,\sigma_{x} ^2, \sigmacon, \phi, \Sigmab \sim \Normal{ \mu_0^{\beta}, \Sigmab \left( \frac{\sigmacon}{\sigma_x ^2} + \psi^2\right) (1-\phi^2)} \ , \,\,
 \Sigmab:= \frac{\Rsquare}{1 - \Rsquare} \sim  \Betapr{\aR, \bR} \ .
\end{eqnarray}
Note that the prior variance
\begin{align} \label{eq:prior_Sigma_0_beta}
    \Sigma_0^{\beta} (\phi,\psi, \sigma_x ^2,\sigmacon ) := \Sigmab \left( \frac{\sigmacon}{\sigma_x ^2} + \psi^2\right)  (1-\phi^2)
\end{align}
depends on several model parameters in a complex manner which needs to be addressed during estimation.
Prior~(\ref{eq:prior_R2D2}) extends the prior obtained in \citet{Wachter_2015} by replacing a fixed 
$\Sigmab$ by a random hyperparameter following a Beta prime distribution, $\Sigmab \sim  \Betapr{\aR, \bR}$. 

The hyperparameters  $\aR$ and $\bR$ have to be chosen with care. 
Note that $\aR=\bR=1$ leads to a uniform prior for $\Rsquare$ and to a Lasso-type marginal prior for $\beta$. For $\aR=\bR= 1/2$, the prior reduces to a horseshoe prior for $\beta$ \citep[see also][]{Cadonna_2020}.

Our simulation study performs Bayesian testing of the null hypothesis
$\mathbb{H}_{0}: \ \beta=0$ against the two-sided alternative $\mathbb{H}_{1} : \ \beta \not=0$ (more details are provided in Section~\ref{sec:simulation}) and in this context we observe that
the choice of $\aR$ is quite influential.
When working with $\bR = 0.5$ and $\aR = 1 / \left( T^{\bR/2}\log T \right) $ as proposed in \citet{Zhang_2020_r2_d2}, the null hypothesis is easily rejected, even if it is true. For the rather small sample size $T=100$ we are interested in, $\aR =  0.07$ and the prior for $\beta$ has a very pronounced spike at zero, causing the Savage-density estimator of the Bayes factor (see Equation~\eqref{eq:savage_density_ratio}) to reject a true null hypothesis too often. We claim that \citet{Zhang_2020_r2_d2} is designed for machine learning applications with a high number of parameters, while in our case the number of parameters is very small and shrinking toward zero is too strong. When working with $\bR = 1$ and $\aR = 0.1$, we observe oversizing, while the Bayesian test rejects a wrong null quite often compared to RBE, if the alternative is true. For the larger value $\aR = 0.5$, we reject a true null hypothesis in approximately 5\% of our simulation runs, while the performance of the test deteriorates, if the alternative is true. Therefore, to learn a suitable value of the parameter $\aR$ from the data, we introduce a hyperprior on  $\aR$. In particular, we apply a Bernoulli distribution, where 
$\aR = \aRb = 0.5$ with probability
$p_{\aR} \in (0,1)$ and
$\aR = \aRu = 0.1$ with probability
$1-p_{\aR}$ and $\bR$ is fixed.

\subsection{MCMC estimation} \label{sec:mcmc}

Assume that realizations of
$\mathbb{D}_T := \{ x_0, \mathbf{x}, \mathbf{y} \}$
from model \eqref{eq:model_1} are available.
Due to the  specific  design of the prior $p(\bm{\theta}) $ introduced in Section~\ref{sec:priors}, the MCMC algorithm applied in this paper to sample from the posterior $p(\bm{\theta}| \mathbb{D}_T )$
required substantial modifications of the Gibbs sampler developed in \citet{Lopes_2014}.
First of all, the prior scale $\Sigma_0^\beta (\phi,\psi, \sigma_x^2, \sigmacon)  $ of our main parameter of interest, $\beta$, 
as defined in  \eqref{eq:prior_Sigma_0_beta}, depends on the regression coefficients $\psi$ and $\phi$ as well as the variances $\sigma_x^2$ and  $\sigmacon$.  As a consequence, most  conditional  posterior distributions no longer belong to well-known distribution families. Furthermore, we are working with the unconditional likelihood (\ref{eq:exact_likelihood}). Nevertheless, in the spirit of \citet{Kastner_2014}, straightforward 
{Metropolis-Hastings (MH)}
algorithms can be applied, where the conditional likelihood serves as a proposal density and the acceptance rate only involves prior ratios. Details are provided in Algorithm~\ref{Algo1}.

\begin{alg}[\textbf{MCMC estimation}] \label{Algo1}
 Choose starting values for $\phi, \sigma_x^2, \psi, \sigmacon,  \Sigmab$ and $a_0^R$. Iterate $M_0 + M_1$ times through the following steps: 
  \begin{itemize} 
 \item[(1)] Sample from $p(\alpY, \beta|  \phi, \psi , \sigma_x^2,\sigmacon, \mathbb{D}_T) $;
\item[(2)] sample from $p(\alpX , \phi|\alpY, \beta, \psi, {\sigma}^{2}_x,\sigmacon, \mathbb{D}_T)$; 
 \item[(3)] sample from $p(\psi|\alpX , \phi,\alpY, \beta, {\sigma}^{2}_x,\sigmacon,\mathbb{D}_T)$;
 \item[(4)] sample from $p(\sigma_x^2|\alpX , \phi, \psi,\beta, \sigmacon, \mathbb{D}_T)$; 
 \item[(5)] sample from $p(\sigmacon|\alpY , \phi, \psi,\beta, \sigma_x^2, \mathbb{D}_T)$;
 \item[(6)] sample from $p(\Sigmab|\phi, \psi,\beta, \sigma_x^2, \sigmacon, a_0^R )$;
 \item[(7)] sample from $p(\aR|\Sigmab)$.
   \end{itemize}
  Discard the first $M_0$ draws as burn-in. Perform thinning by keeping one in $C$ draws   and discarding all others. This results  in $M= \lfloor  M_1/C \rfloor$ MCMC draws.
\end{alg}

To sample the regression parameters
$(\alpY, \beta) $ and $(\alpX,\phi)$ in Step~1 and 2 of Algorithm~\ref{Algo1}, we move
to representation \eqref{eq:model_1}. 
Given the current MCMC draws of $\phi$, $\sigma_x^2$, $\sigmacon$ and $\psi$,
we adjust the scale  $\Sigma_0^\beta (\phi,\sigma_x^2, \sigmacon,\psi)  $ in the prior distribution
$\alpY, \beta|\phi, \psi , \sigma_x^2,\sigmacon \sim \Normal{\mum_{0}, \Sigmam_{0}}$, where
    \begin{align*}
    \mum_{0} = \begin{pmatrix} \mu_0^{\alpY} \\ \mu_0^{\beta} \end{pmatrix},
    \qquad \Sigmam_{0}  =
    \begin{pmatrix} 
        \Sigma_0^{\alpY} & 0 \\ 0 & \Sigma_0^{\beta} (\phi,\sigma_x^2, \sigmacon,\psi) 
    \end{pmatrix},
    \end{align*}
and define the variances
$\sigma_y^2=\sigmacon + \sigma_x^2 \psi^2$ and 
$\sigma_{xy}=\psi \sigma_{x}^2$ in system  \eqref{eq:model_1}.     
The posterior $\alpY, \beta|\phi, \psi , \sigma_x^2,\sigmacon, \mathbb{D}_T \sim \Normal{\mum_{T}, \Sigmam_{T}}$ is then a bivariate Gaussian distribution with 
\begin{align}\label{eq:mcmc_full_posterior_beta}
    \Sigmam_{T} &= (\Xbetamat ^\top \Xbetamat /{\sigma}_y^{2}+\Sigmam_{0}^{-1})^{-1} \ , \nonumber \\
    \mum_{T} &= \Sigmam_{T} (\Xbetamat ^\top \ym /\sigma_y^{2} + \Sigmam_{0}^{-1}\mum_{0}) \ ,
\end{align}
where $\Xbetamat$ is a $T \times 2$ matrix with the $t$-th row given by $\Xbeta_{t}= (1,\ x_{t-1})$. The errors $\epsilon_t^{y}$ obtained from the return equation of model~(\ref{eq:model_1}) for all $t=1, \ldots, T$ under the current draws of $\alpY$ and $\beta$ are then used to sample $\alpX$ and $\phi$. Using 
\begin{align*}
    \epsilon_t^{x}| \epsilon_t^{y} \sim \mathcal{N}
\left( \frac{\sigma_{xy}}{\sigma_{y}^2}
\epsilon_t^y, \sigma_{x}^2 - \frac{\sigma_{xy}^2}{\sigma_{y}^2}
\right),
\end{align*} 
we obtain the following representation for the dividend-price ratio equation of model~(\ref{eq:model_1}):
\begin{align}
		x_{t} - \tilde{u}_t &= \alpX + \phi x_{t-1} + \tilde{\epsilon}_{t}^x, \qquad \tilde{\epsilon}_{t}^x \sim \mathcal{N}(0,\tilde{\sigma}^2_{x}), \label{eq:step_1_1} \\
		y_t &= \alpY + \beta x_{t-1} + \epsilon_t^y \label{eq:step_1_2}, 
	\end{align}
where $\tilde{u}_t$ and $\tilde{\sigma}^2_{x}$ can be expressed in terms of the parameters $\psi$, $\sigma_x^2$ and $\sigmacon$:
	\begin{align}\label{eq:step_1_3}
		\tilde{u}_t=\frac{\sigma_{xy}}{\sigma_{y}^2}\epsilon_t^y = \frac{\sigma_x^{2}\psi}{\sigmacon +\sigma_x^2\psi^2} \epsilon_t^y \ ,
  \qquad  \tilde{\sigma}^2_x=
   \sigma_{x}^2 - \frac{\sigma_{xy}^2}{\sigma_{y}^2}=
 \frac{\sigma_x^2\sigmacon}{\sigmacon+\sigma_x^2\psi^2}.
	\end{align}
The posterior of $\alpX$ and $\phi$ is derived from 
the `bias corrected' dividend-price ratio equation~\eqref{eq:step_1_1}, where  we condition on the current errors $\epsilon_t^y$ of the return equation~\eqref{eq:step_1_2}. 
{Appendix~\ref{app:mcmc_details_MH} discusses further details on sampling from the conditional posterior $p(\alpX , \phi|\alpY, \beta, \psi, {\sigma}^{2}_x,\sigmacon, \mathbb{D}_T)$.}

The remaining parameters are sampled in representation~\eqref{eq:model_3}. Since we did not change the parameters defining $\Sigmam$, the errors $\epsilon_t^{x}$ 
and  $\epsilon_t^{y}$ obtained from model (\ref{eq:model_1}) for all $t=1, \ldots, T$ under the current draws of $(\alpX,\phi,
\alpY,\beta)$ define sufficient statistics
to sample $\psi$ from the return equation of representation~\eqref{eq:model_3}. The posterior 
   $p(\psi|\alpX , \phi,\alpY, \beta, {\sigma}^{2}_x,\sigmacon,\mathbb{D}_T) \propto
    p(\mathbf{y}| \psi,\alpY, \beta, \mathbf{x}) p(\psi)  p(\beta| \psi) $
is non-Gaussian, since the prior $\beta| \psi \sim \Normal{0,\Sigma_0^\beta (\psi,\cdot)}$ depends on $\psi$ in a non-conjugate manner. However, a Gaussian density is obtained from combining the likelihood 
$ p(\mathbf{y}| \psi,\alpY, \beta, \mathbf{x} ) $ with the Gaussian prior $p(\psi)$:
\begin{align}\label{eq:posterior_psi}
  &q(\psi |\cdot)  \sim \Normal{\mu_T^\psi, \Sigma_T^\psi},  \\
  &\Sigma_T^\psi =  \left( (\Sigma_0^\psi) ^{-1}  + \frac{1}{\sigmacon} \sum_{t=1}^T (\epsilon_t^{x})^2  \right)^{-1},  \quad \mu_T^\psi = \Sigma_T^\psi \left(  (\Sigma_0^\psi) ^{-1} \mu_0^\psi +   \frac{1}{\sigmacon} \sum_{t=1}^T \epsilon_t^{x} \epsilon_t^{y} \right). \nonumber 
\end{align}
Taking this Gaussian density as proposal density for a new value $\psi \new $, the term $A$ in the acceptance rate  $\min\{1,A\}$ is simply the prior ratio:
 \begin{align*}
	A= \frac{ p \left( \beta| \Sigma_0^\beta ( \psi \new,\cdot) \right)}
	{p \left( \beta| \Sigma_0^\beta (\psi ,\cdot) \right)}.
\end{align*}
The complex dependence of $\Sigma_0^\beta$ on both innovation variances $\sigma_x^2$ and $\sigmacon$ is addressed similarly. The prior $\beta| \sigma_x^2, \cdot \sim \Normal{0,\Sigma_0^\beta (\sigma_x^2, \cdot)}$ is combined with the inverse gamma posterior 
\begin{align}  \label{eq:posterior_sig_x}
    &q(\sigma_{x}^{2}| \cdot) \sim \Gammainv{\nuX + \frac{T}{2}, S_0^x + \frac{1}{2}\sum_{t=1}^T (\epsilon_t^{x})^2},
\end{align}
that is obtained from combining the likelihood 
$p \left(\mathbf{x} , x_0| \alpX, \phi, \sigma_{x}^2 \right) $ 
with the inverse gamma prior of $\sigma_x^2$. An MH-step  with the inverse gamma distribution (\ref{eq:posterior_sig_x}) serving as a proposal density for $(\sigma_{x}^{2}) \new$ is performed, where the acceptance rate  $\min\{1,A\}$ simply depends on the prior ratio:
 \begin{align*} 
	A= \frac{p \left(\beta \left| \Sigma_0^\beta \left((\sigma_{x}^{2}) \new, \cdot\right) \right. \right)} {p\left(\beta \left| \Sigma_0^\beta \left( \sigma_{x}^{2} 
, \cdot \right) \right. \right)}.
\end{align*}
Similarly, the posterior of $ \sigmacon$ is no longer inverse gamma. The prior $\beta| \sigmacon, \cdot \sim \Normal{0,\Sigma_0^\beta (\sigmacon, \cdot)}$ is  combined with the inverse gamma posterior
\begin{align} \label{eq:posterior_sig_tilde_y}
    & q(\sigmacon | \cdot) \sim \Gammainv{\nuY + \frac{T}{2}, \SY + \frac{1}{2}\sum_{t=1}^T (\tilde{\epsilon}_t^{y})^2},
\end{align}
that is obtained from combining the likelihood 
$p \left(\mathbf{y} | \alpY, \psi, \sigmacon, \sigma_{x}^2, \mathbf{x} \right) $ 
with the inverse gamma prior of $\sigmacon$. Again, a MH-step  with the inverse gamma distribution (\ref{eq:posterior_sig_tilde_y}) serving as a proposal density for $(\sigmacon) \new$ yields an acceptance rate  $\min\{1,A\}$ depending only on the prior ratio:
\begin{align*}
    A= \frac{p \left( \beta| \Sigma_0^\beta \left( (\sigmacon) \new , \cdot\right) \right)}
    {p \left( \beta| \Sigma_0^\beta  \left(\sigmacon , \cdot \right)  \right)}.
\end{align*}
Finally, in  {Steps}~6 and 7, we update the hyperparameters
$\Sigmab$ and $\aR$ defining the prior on $\beta$.
Using an alternative representation of the beta prime distribution\footnote{$X \sim \Betapr{a ,b}  \quad \Leftrightarrow \quad X| Z \sim \Gammainv{b, Z}, \quad Z  \sim \Gammad{a,  1}$} we obtain the following hierarchical representation of the prior for $\beta$ given in Equation~\eqref{eq:prior_R2D2}:
\begin{align} \nonumber 
   \beta | \Sigmab, \cdot \sim \Normal{ 0, \Sigmab  \left( \frac{\sigmacon}{\sigma_x ^2} + \psi^2\right) (1-\phi^2)}, \,\, \Sigmab | Z_\beta \sim \Gammainv{\bR, Z_\beta},  \,\,  Z_\beta|\aR  \sim \Gammad{\aR ,  1}.
\end{align}
Given $\Sigmab$, we first impute $Z_\beta$ from $Z_\beta|\Sigmab, \aR \sim \Gammad{\aR + \bR,  1 + \frac{1}{\Sigmab}}$. Then, we sample from $ \Sigmab $ conditional on $Z_\beta$ and all remaining parameters:
\begin{align}  \label{eq:posterior_sigma_eta}
    \Sigmab| \phi, \psi,\beta, \sigma_x^2, \sigmacon, a_0^R     \sim \Gammainv{\bR + \frac{1}{2}, Z_\beta + S_\beta }  , \quad S_\beta=\frac{\beta^2}{2 (1-\phi^2)}  \left( \frac{\sigmacon}{\sigma_x ^2} + \psi^2\right)^{-1},
\end{align}
for details see Appendix \ref{app:mcmc_details_Sigmab}.
Finally, since $\aR$ depends on {$\Sigmab$} in a complex manner we apply the MH to sample from the posterior distribution $p(\aR|\Sigmab)$, where $A$ in the acceptance rate  $\min\{1,A\}$ simplifies to
\begin{align*}
    A = \frac{p \left({\Sigmab}|\aRnew \right)}{p \left(  {\Sigmab}|\aR \right)}.
\end{align*}
Details are provided in Appendix~\ref{app:mcmc_details_aR}.

\section{Performance in Simulated Data} \label{sec:simulation}

This section investigates the finite sample performance of ordinary least square estimation, reduced-bias estimation  [described in equation~\eqref{eq:RBE} and proposed by \citet{Amihud_2004}], and the Bayesian approach introduced in this paper. We analyse two data-generating processes ({\em DGP 0} and {\em DGP 1}), where the model parameters are chosen similarly to the estimates obtained for the empirical data sets considered in  Section~\ref{sec:financial}. That is, we generate $T = 100$ observations by using model~(\ref{eq:model_1}) with  parameter values $\alpX = -0.15$, $\alpY = 0.6$, $\phi = 0.95$ and $\sigma_x^2 = 0.02$, $\sigma_y^2 = 0.04$, $\sigma_{xy} = -0.02$. The main difference between {\em DGP 0} and {\em DGP 1} lies in the value of the predictive regression coefficient $\beta$. {\em DGP 0} sets $\beta = 0$ which corresponds to the situation of no predictability. {\em DGP 1} fixes $\beta = 0.1$ that resembles weak return predictability. In Appendix~ \ref{app:sim_res_diff_beta} we also analyze the performance of the estimator for 
$\beta \in \{0, 0.025, 0.05, 0.075, 0.1, 0.2\}$. 

Let $\xi$ denote a specific coordinate of the parameter vector ${\bm \theta}$ and let $\widehat{\xi}^{\cdot}$ denote a point estimator of  $\xi$.  
The OLS estimator of $\xi$ is abbreviated by $\widehat{\xi}^{OLS}$, with $SE \left( \widehat{\xi}^{OLS} \right)$ denoting the corresponding standard error. For reduced-bias estimation, $\widehat{\xi}^{RBE}$ denotes the point estimate and $SE \left( \widehat{\xi}^{RBE} \right)$ the standard error obtained in \citet{Amihud_2004}. For the Bayesian approach (BAY),  we abbreviate the Bayesian point estimator (in this paper given by the posterior mean of the MCMC samples after burn-in and thinning) by $\widehat{\xi}^{BAY}$. 

For the Bayesian approach, we choose the following hyperparameters for the priors defined in Section~\ref{sec:priors}.
We specify the prior means for the regression coefficients $\alpY$ and $\psi$ to be 
$\mu_0^{\alpY} = \mu_0^{\psi} = 0$ and the prior variances to be $\Sigma_0^{\alpY} = \Sigma_0^{\psi} = 10$. We further specify the prior mean $\mualpX=1/(T+1)\sum_{t=0}^T x_t$ and variance $\SalpX=0.2$ for $\alpX$ that are based on the observed data. Next, we specify the prior shapes for the variances $\sigma_{x}^2$ and $\sigmacon$ to be $\nuY = 2.5, \nuX = 4$ and prior scales to be $\SY = 0.03$ and $S_0^x = 0.06$. 
Finally, to define the prior on $\beta$, we assume $\mu_0^{\beta} = 0$.  For the parameter $\aR$  we apply a Bernoulli distribution, where 
$\aR = \aRb = 0.5$ with probability 
$p_{\aR } = 0.5$ and 
$\aR  = \aRu = 0.1$ with probability 
$ 1-p_{\aR} = 0.5$, while $\bR = 1$.

To obtain draws  ${\bm \theta}^{(m)}$, $m=1,\dots,M$ from the corresponding posterior distribution, we use the MCMC sampler described in Algorithm~\ref{Algo1} in Section~\ref{sec:mcmc}.
The number of original MCMC draws equals $M_0 + M_1 = 100\,000$, {discarding the first $M_0 = 10\,000$ draws as burn-in steps}. We perform thinning by taking one in forty-five MCMC draws ({$C = 45$}) and discarding all others -- resulting in $M=2\,000$ MCMC draws that we use for further inference.

For each data generating process, we simulate $i=1,\dots,n = 1\,000$ data sets $\mathbb{D}_{T,i}$ and keep only those, where the MCMC sampler exhibits a satisfactory convergence. We identify proper convergence by computing the Effective Sample Size (ESS) and analysing the trace plots for the thinned sample, see Appendix~\ref{app:mcmc_details_convergence} for more details. The MCMC sampler exhibited  in general satisfactory convergence properties and we kept $n_d = 993$ data sets under {\em DGP 0} and $n_d = 956$ data sets under {\em DGP 1}. 
 
These $n_d$ data sets  are used for further evaluation of all three estimation methods and the corresponding point estimates are denoted  by $\widehat{\xi}_i^{\cdot}$, $i=1,\dots,n_d$. 
For the Bayesian approach, $\widehat{\xi}_i^{BAY}$ is given as the mean of (thinned)  posterior draws for a specific data set $i$.
To evaluate the quality of the estimator $\widehat{\xi}^{\cdot}$, we calculate [estimates of] the bias (denoted $B$), the standard deviation of the point estimator ($\sigma$), the root mean square error (RMSE) and the mean absolute error (MAE),
where $\xi$ is the true value of the parameter:
\begin{align}
    B(\widehat{\xi}, \xi) &:= \frac{1}{n_d}\sum_{i=1}^{n_d}\left( \widehat{\xi}_i - \xi \right)\ , \ 
    &&\sigma(\widehat{\xi}) := \sqrt{\frac{1}{n_d}\sum_{i=1}^{n_d} \left( \widehat{\xi}_i - \frac{1}{n_d}\sum_{i=1}^{n_d} \widehat{\xi}_i \right)^2} \ ,  \nonumber \\ 
	RMSE(\widehat{\xi}, \xi) &:= \sqrt{\frac{1}{n_d}\sum_{i=1}^{n_d} \left( \widehat{\xi}_i - \xi \right)^2} \ , \	&&
  MAE \left( \widehat{\xi}, \xi \right) := \frac{1}{n_d}\sum_{i=1}^{n_d}\left| \widehat{\xi}_i - \xi \right|  \ .  \label{eq:MAE}
\end{align}  
Tables~\ref{tbl:phi_sim_data_rmse} and \ref{tbl:beta_sim_data_rmse} present summary statistics for the $\widehat{\phi}^{\cdot}$ and $\widehat{\beta}^{\cdot}$, for data generated by {\em DGP 0} and {\em DGP 1}, respectively. For the parameter $\phi$, the RBE estimator achieves the smallest bias, followed by our Bayesian estimator (BAY) and the OLS estimator. The Bayesian estimator has the smallest standard deviation, MAE and RMSE.

\begin{table}[h!] \centering 
	\caption{Measures of the estimation quality for $\widehat{\phi}$ (multiplied by 100): Bias $B(\widehat{\phi}, \phi)$, Sample standard deviation of the point estimates $\sigma(\widehat{\phi})$, MAE, RMSE. Three estimation methods: OLS, RBE and BAY. {\em DGP 0} sets $\beta = 0$, whereas {\em DGP 1} fixes $\beta = 0.1$.} 
	\label{tbl:phi_sim_data_rmse} 
	\begin{tabular}{lcccc} 
 \toprule
		Method & $B(\widehat{\phi}, \phi)$ & $\sigma(\widehat{\phi}$) & MAE($\widehat{\phi}, \phi)$ & RMSE($\widehat{\phi}, \phi)$ \\ \midrule
		\multicolumn{5}{c}{\em DGP 0}\\
		\midrule
OLS & -4.37 & 5.01 & 4.88 & 6.65  \\ 
RBE & -0.50 & 5.16 & 3.91 & 5.19  \\ 
BAY & -1.10 & 3.41 & 2.39 & 3.58  \\
		\midrule
		\multicolumn{5}{c}{\em DGP 1}\\
		\midrule
OLS & -4.44 & 5.05 & 4.93 & 6.73  \\ 
RBE & -0.57 & 5.21 & 3.93 & 5.24  \\ 
BAY & -0.80 & 3.74 & 2.59 & 3.83  \\ 
\bottomrule
	\end{tabular} 
\end{table}

\begin{table}[ht] \centering 
	\caption{Measures of the estimation quality for $\widehat{\beta}$ (multiplied by 100): Bias, Sample standard deviation of the point estimates $\sigma(\widehat{\beta})$, MAE, RMSE, estimated False Positive ($\widehat{FP}$) and estimated False Negative rates ($\widehat{FN}$). Three estimation methods: OLS, RBE and BAY. {\em DGP 0} sets $\beta = 0$, whereas {\em DGP 1} fixes $\beta = 0.1$.} 
	\label{tbl:beta_sim_data_rmse} 
	\begin{tabular}{lccccc} 
 \toprule
		Method & $B(\widehat{\beta}, \beta$) & $\sigma(\widehat{\beta}$) & MAE($\widehat{\beta}, \beta$) & RMSE($\widehat{\beta}, \beta$) & -  \\ \midrule
		\multicolumn{5}{c}{\em DGP 0} & $\widehat{FP}$ \\
		\midrule
OLS & 4.33 & 6.56 & 5.77 & 7.86 & 8.14 \\ 
RBE & 0.47 & 6.69 & 5.10 & 6.71 & 7.21 \\ 
BAY: $\aR$ random & 1.94 & 3.93 & 2.55 & 4.38 & 6.14 \\ 
\midrule
BAY: $\aR = 0.1$ fixed & 1.32 & 2.69 & 1.62 & 2.99 & 7.36 \\ 
BAY: $\aR = 0.5$ fixed & 3.32 & 5.09 & 4.17 & 6.07 & 4.52 \\ 
		\midrule
		\multicolumn{5}{c}{\em DGP 1} & $\widehat{FN}$\\
		\midrule
OLS & 4.46 & 6.67 & 5.89 & 8.02 & 28.3 \\ 
RBE & 0.60 & 6.80 & 5.20 & 6.83 & 63.27 \\ 
BAY: $\aR$ random & 0.27 & 6.83 & 5.16 & 6.84 & 36.19 \\ 
\midrule
BAY: $\aR = 0.1$ fixed & -2.11 & 6.67 & 5.78 & 6.99 & 34.91 \\ 
BAY: $\aR = 0.5$ fixed & 1.58 & 6.2 & 4.82 & 6.39 & 52.98 \\ 
\bottomrule
	\end{tabular}
\end{table}

For the parameter $\beta$, OLS estimation results in a substantial bias as can be expected by the results obtained in \citet{Stambaugh_1999}. Due to a negative bias of the OLS estimate of the parameter $\phi$ and a negative estimate of $\sigma_{xy}$, we observe an upward bias with $\widehat{\beta}^{OLS}$. The biases for $\widehat{\beta}^{RBE}$
and $\widehat{\beta}^{BAY}$ are much smaller, where $B(\widehat{\beta}^{BAY},\beta) 
> B(\widehat{\beta}^{RBE},\beta)$ for {\em DGP 0}, while 
$B(\widehat{\beta}^{BAY},\beta) 
< B(\widehat{\beta}^{RBE},\beta)$ for {\em DGP 1}. 
Regarding MAE, the Bayesian estimator outperforms the RBE and the OLS estimator. For {\em DGP 0} the RMSE is smaller with BAY compared to the other estimators, while for {\em DGP 1} RMSE with the RBE is slightly smaller than with BAY. The largest RSME is observed with OLS estimation.

This result is also illustrated in Figure~\ref{fig:plot.beta}, where the sampling distributions of {$\widehat{\beta}^{OLS}$, 
$\widehat{\beta}^{RBE}$ and $\widehat{\beta}^{BAY}$ are provided.} As expected, the sampling distributions of $\widehat{\beta}^{.}$  under {\em DGP 0} {(left panel in Figure~\ref{fig:plot.beta})}  is centered around zero for all estimation techniques. 
The Bayesian point estimator is the most efficient, with a sampling distribution that is more concentrated at values near zero. That supports the evidence from Table~\ref{tbl:beta_sim_data_rmse} that the Bayesian estimate has the lowest RMSE and MAE. For {\em DGP 1} (right panel in Figure~\ref{fig:plot.beta}), RBE and the Bayesian estimate have similar performance and both outperform the OLS in terms of RMSE and MAE.

\textsc{\begin{figure}[t]
		\includegraphics[width = 15cm]{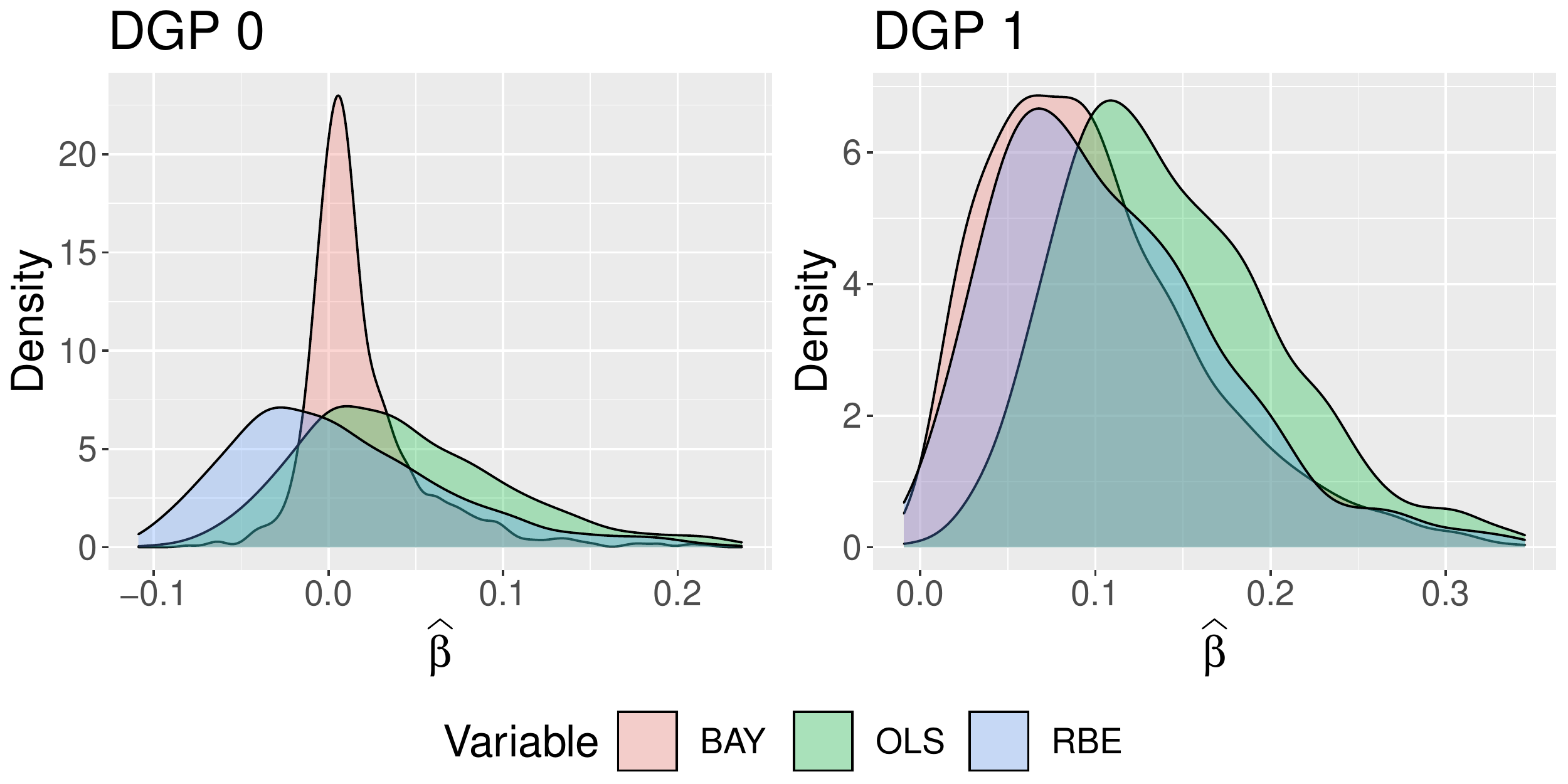}
		\centering
		\caption{Sampling distributions of the point estimator $\widehat{\beta}$ for the Bayesian estimator (red) and for the frequentist estimators OLS (green) and RBE (blue). The left panel is the result for {\em DGP 0} and the right panel is the result for {\em DGP 1}.}
		\label{fig:plot.beta}
\end{figure}}

Next, we discuss the performance of the various approaches when testing the null hypothesis $\mathbb{H}_0: \ \beta=0$ against the two-sided alternative $\mathbb{H}_1: \ \beta \not=0$ for the $n_d$ data sets under the two simulation settings {\em DGP 0} and {\em DGP 1}.
For OLS and RBE estimation we perform $t$-tests with a significance level of 5\%. In a Bayesian setting, we opt for the null hypothesis, if the \textit{Bayesian factor} ($BF_{01}$) of $\mathbb{H}_{0}$ versus $\mathbb{H}_{1}$ is larger than 1. The Bayes factor can be expressed as the ratio of the posterior over the prior ordinate  at zero obtained under $\mathbb{H}_1$, i.e.:
\begin{align}
	BF_{01} := \frac{p(\beta = 0|\mathbb{H}_{1},\mathbb{D}_T )}{p(\beta = 0|\mathbb{H}_{1})}  \ , \label{eq:savage_density_ratio}
\end{align}
where the fraction on the right-hand side in 
(\ref{eq:savage_density_ratio}) is called the 
{\em Savage-Dickey Density Ratio} \citep[see][and \citet{wagenmakers2010bayesian} for a detailed derivation.]{Dickey_1970}. 
The {estimated Bayes factor ($\widehat{BF}_{01}$)} is determined for each data set through an approximation of the  Savage-Dickey Density Ratio based on the MCMC draws of $\beta$. Details are provided in Appendix~\ref{app:simulation_details_BF}.

To evaluate this Bayesian procedure, we consider the false positive rate (that is, the probability a statistical test rejects a true null hypothesis; abbreviated by ${{FP}}$) and the false negative rate (${FN}$,  that is, the probability a test rejects a true alternative hypothesis). In this article, the {\em estimated False Positive Rate} (${\widehat{FP}}$; observed size with the OLS and RBE estimator) is the relative frequency of rejecting the true null hypothesis $\beta=0$, while the {\em estimated False Negative Rate} (${\widehat{FN}}$; one minus the observed power with OLS and RBE estimation) is the relative frequency of rejecting a true alternative-hypothesis $\beta \not=0$.

Under a theoretical significance level of 5\%, we observe in Table~\ref{tbl:beta_sim_data_rmse} quite substantial oversizing for the frequentist tests when the null hypothesis is true. For the Bayesian procedure, the priors are chosen such that ${\widehat{FP}}$ becomes close to the theoretical significance level of 5\%. In the case of $\beta \not= 0$, we consider ${\widehat{FN}}$ (one minus the observed power). In this case, our Bayesian procedure has a much smaller ${\widehat{FN}}$ rate compared to RBE. That is, the observed power of RBE is poor and the Bayesian estimate dominates the RBE in terms of ${\widehat{FP}}$ as well as ${\widehat{FN}}$. Due to a substantial bias and small standard errors, the null hypothesis is rejected more often with OLS estimation, which -- on the other hand -- results in high observed power (${\widehat{FN}}$ in Bayesian terminology).

Table~\ref{tbl:beta_sim_data_rmse} [and in more details
Appendix \ref{app:sim_res_diff_beta}] presents the estimated false positive (${\widehat{FP}}$) and false negative (${\widehat{FN}}$) rates based on the estimated Bayes 
{factors $\widehat{BF}_{01,i}$, $i=1,\dots,n_d$,} with $\aR$ being fixed at $0.1$ or $0.5$ or being random, with a uniform distribution over these two values. Here we observe an interesting trade-off between estimation and testing errors. Under {\em DGP 0}, choosing the smaller value of $\aR$ yields smaller MAE and RMSE, but at the same time a higher ${\widehat{FP}}$ rate. Under {\em DGP 1}, choosing the larger value of $\aR$ yields smaller MAE and RMSE, but a higher ${\widehat{FN}}$ rate. The hyperprior on $\aR$ improves this situation by finding a comprise between these two extremes. Figure~\ref{fig:aR_posterior} provides an intuitive reason why this improvement takes place. Figure~\ref{fig:aR_posterior} depicts the sampling distribution of the posterior mean $\widehat{a}_0^R$, both for {\em DGP 0} (red) and for {\em DGP 1} (blue). While $\aR$ takes only two values $0.1$ or $0.5$, the posterior mean 
of ${a_0^R}$ is a continuous variable  that lies between $0.1$ and $0.5$. For each data set $\mathbb{D}_{T,i}$, the posterior mean of 
$a_0^R$ is provided by
$\widehat{a}_{0,i}^R := \frac{1}{M}
\sum_{m=1}^M a_{0,i}^{R,(m)}$, $m=1,\dots,M$,
where $a_{0,i}^{R,(m)}$ are MCMC draws of $a_0^{R}$ for the corresponding data set $\mathbb{D}_{T,i}$. Under {\em DGP 0} we learn that we have no predictability and hence $\widehat{a}_0^R$ (and the corresponding coefficient of determination) is smaller, imposing on average larger shrinkage ex-ante, compared to the {\em DGP 1}, where the posterior mass of $\widehat{a}_0^R$ is larger.

\textsc{\begin{figure}[t]
		\includegraphics[width = 15cm]{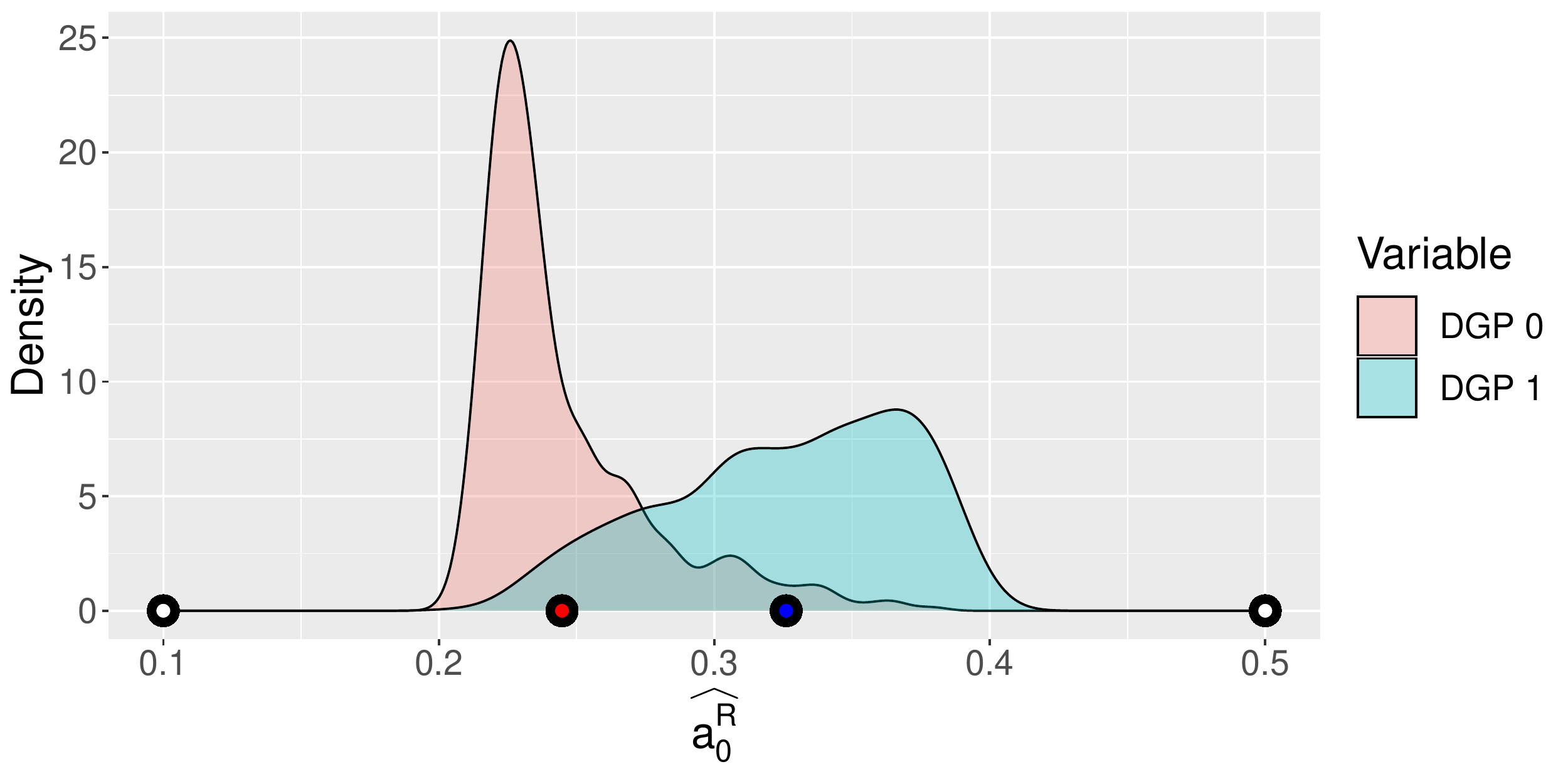}
		\centering
		\caption{Sampling distribution of the posterior mean $\widehat{a}_0^R$ based on posterior means $\widehat{a}_{0,i}^R$, $i=1,\dots,n_d$ obtained from the thinned draws $(a_0^R)_i^{(m)}$, $m=1,\dots,2\,000$ for the Bayesian estimator for {\em DGP 0} (red) and for {\em DGP 1} (blue). White dots represent the prior values for $\aR$ with $\aRu = 0.1$,  $\aRb = 0.5$ and two filled dots represent the average of the posterior means $\widehat{a}_{0,i}^R$ for {\em DGP 0} (red) and for {\em DGP 1} (blue).}
		\label{fig:aR_posterior}
\end{figure}}

To further explore the intuition behind the testing procedure using the Bayes Factor
{Figure~\ref{fig:beta_avg_posterior} presents the 5/50/95\% quantiles of estimated marginal posterior distributions $\widehat{p}\left({\beta}|\mathbb{D}_{T,i} \right)$,
$i=1,\dots,n_d$,
by applying the kernel density estimator implemented in {\tt R}, using default bandwidth choice}.  We observe that for {\em DGP 0} we have a very spiky distribution for the median
of $\widehat{p}\left({\beta}|\mathbb{D}_{T,i} \right)$, with the spike of the shrinkage prior at $0$ being amplified by the data observed. For {\em DGP 1} the median of 
$\widehat{p}\left({\beta}|\mathbb{D}_{T,i} \right)$
also has a peak at $\beta$ equal to zero. We claim that this effect is mainly caused by our shrinkage prior. 
With a true value of $\beta=0.1$ and $T=100$, the data is sufficiently informative to have a further, but smaller peak at $0.1$. Appendix~\ref{app:sim_res_diff_beta_and_T_len} demonstrates that for values of $\beta$ equal to $0.2$ 
{(or larger)} and a number of observations $T \geq 1\,000$ the likelihood overcomes the effect of the prior.

\textsc{\begin{figure}[t]
		\includegraphics[width = 15cm]{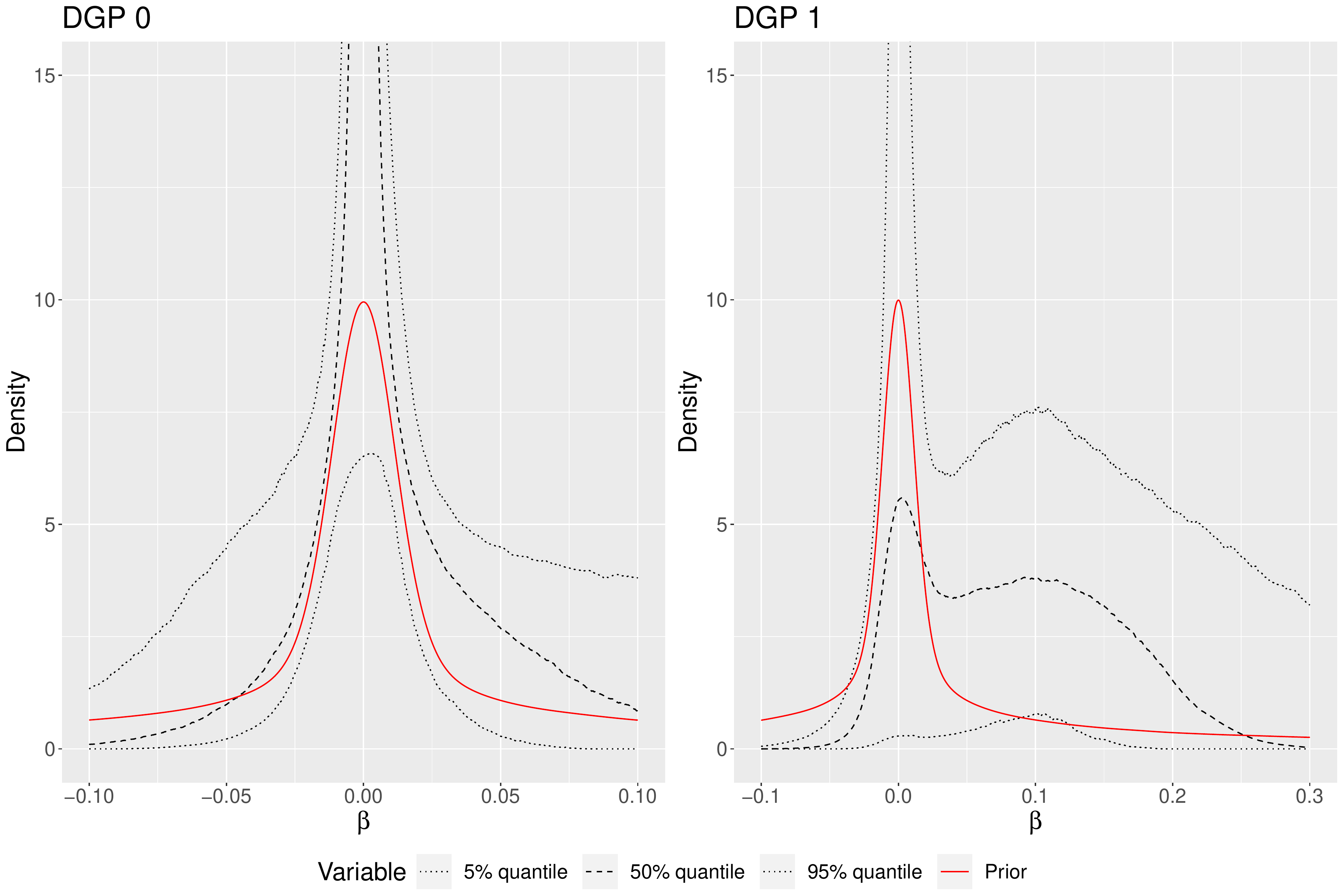}
		\centering
		\caption{5/50/95\% quantiles (dashed lines) of the estimated marginal posterior distributions $\widehat{p}\left({\beta}|\mathbb{D}_{T,i} \right)$ by the kernel density estimator implemented in {\tt R}, using default bandwidth choice using the thinned draws $\beta^{(m)}_i$, $m=1,\dots,2\,000$ for $i=1,\dots,n_d$. The solid red line shows the prior density. The left panel is the result for {\em DGP 0} and the right panel is the result for {\em DGP 1}.
  }
\label{fig:beta_avg_posterior}
\end{figure}}

In Table~\ref{tbl:beta_sim_data_rmse} we observe that 
for our Bayesian test
the null hypothesis $\beta=0$ was falsely rejected in approximately 6\% of the simulation runs for {\em DGP 0}, while with {\em DGP 1} we observe a false rejection rate of approximately 36\%. Recall that the Bayes factor $BF_{01}$ is obtained by the {\em Savage-Dickey Density Ratio}~(\ref{eq:savage_density_ratio}). In the left panel of Figure~\ref{fig:beta_avg_posterior} the median of the posteriors $\widehat{p}(\beta =0 | \mathbb{D}_{T, i} )$, $i=1,\dots,n_d$, is far above the prior $p(\beta =0 )$ resulting in no rejection of the true null hypothesis with {\em DGP 0} in approximately 6\% of the simulation runs. In the right panel of Figure~\ref{fig:beta_avg_posterior} we observe the results for {\em DGP 1}, where the median of the posteriors $\widehat{p}(\beta =0 | \mathbb{D}_{T, i} )$, $i=1,\dots,n_d$, is below
$p(\beta =0 )$. The fraction of samples where 
$\widehat{BF}_{01} > 1$ approximately corresponds to the false negative rate presented in Table~\ref{tbl:beta_sim_data_rmse}. The calculation of  $\widehat{BF}_{01}$ in Table~\ref{tbl:beta_sim_data_rmse}, as 
explained in Appendix~\ref{app:simulation_details_BF}, 
slightly differs from the eyeballing density comparison in Figure~\ref{fig:beta_avg_posterior}.

\section{Application to the Financial Data} \label{sec:financial}

Our empirical analysis is based on two datasets. The first dataset (\textit{Sample 1}) is based on \citet{Cochrane_2008}\footnote{We thank John Cochrane to provide data at the website \href{https://www.johnhcochrane.com/the-dog-that-didnt-bark}{https://www.johnhcochrane.com/the-dog-that-didnt-bark}.} and 
comprises 78 annual observations of CRSP value-weighted returns from January 1926 until December 2004. Our second dataset 
(\textit{Sample 2)} is a post-war sample from January 1953 until December 2021 containing 68 observations\footnote{We thank Amit Goyal to provide data at the website \href{http://www.hec.unil.ch/agoyal/}{http://www.hec.unil.ch/agoyal/}.}. 
We construct the dividend-price ratio 
{$X^{DP}_{t}$} following the standard procedure in the literature \citep[see, e.g.,][]{Cochrane_2011}, where we denote the dividend-price ratio as
\begin{align}\label{eq:dp_ratio_construction}
	X^{DP}_{t} := \frac{D_{t}}{P_{t}} = \frac{R_{t}}{R^{-}_{t}} - 1,
\end{align}
where {$R^{-}_{t}$} is the CRSP value-weighted returns excluding the dividends and $R_{t}$ is the CRSP value-weighted returns including the dividends defined as
\begin{align*}
    R_{t} := \frac{P_{t} + D_{t}}{P_{t-1}}, \qquad R^{-}_{t} := \frac{P_{t}}{P_{t-1}},
\end{align*}
where $P_t$ is the price at the end of the year $t$ and $D_{t}$ is the sum of dividends during the year $t$. The log dividend-price ratio constructed from Equation~\eqref{eq:dp_ratio_construction} and CRSP value-weighted log return defined as
\begin{align*}
	x_{t} = \log \left(X^{DP}_{t}\right) = \log D_t - \log P_t \ , \qquad y_{t}  = \log(R_{t}).
\end{align*}

For these empirical data sets, we apply model~(\ref{eq:model_1}) and estimate the model parameters by means of ordinary least squares, the reduced-bias estimator of \citet{Amihud_2004}, and the Bayesian approach described in Section~\ref{sec:bayesian_analysis}. The parameters of the prior remain the same as in Section~\ref{sec:simulation}. Our main focus is on the parameter $\beta$, which relates to the question whether asset returns can be predicted by means of the lagged log dividend-price ratio. When applying the Bayesian sampler, we generate {$M_0+M_1=$}  1\,000\,000 MCMC draws, burn-in equals {$M_0=$} 100\,000 draws and thinning equals 
{$C=$} 30, that results in 30\,001 posterior draws. Some trace plots and convergence analysis are provided in Appendix~\ref{app:financial_data_traceplots}.

Table~\ref{tbl:beta_results} presents parameter estimates 
$\widehat{\beta}^{\cdot}$ and results obtained from frequentist and Bayesian testing. For both samples, we observe that the OLS estimates are approximately twice as large as the estimates obtained with RBE and BAY. As already observed in Section~\ref{sec:simulation}, we claim that this effect is mainly caused by the upward bias observed with OLS estimation. The point estimates with BAY and RBE are quite similar in both samples. By considering the tests on the null hypothesis that $\beta=0$ against the two-sided alternative $\beta \not= 0$ for OLS estimation the $p$-values remain $\leq $ 6.2\%, while for the RBE the $p$-values are $\geq$ 19\%. In our simulation study, we observed strong oversizing for OLS estimation and a low empirical power with RBE. By contrast, while the  estimated false positive rates (${\widehat{FP}}$) remained close to 5\% with our Bayesian estimation, the estimated false negative rates (${\widehat{FN}}$) were much smaller than with RBE. Hence, one rationale to apply our Bayesian procedure is that BAY dominates RBE in terms of observed size and power. By using the Bayes factor we observe no predictability for \textit{Sample~1} and weak evidence for predictability with \textit{Sample~2}. 

\begin{table}[H]
	\centering
	\caption{Results for $\beta$: Parameter estimate
 $\widehat{\beta}^{\cdot}$, the standard error $SE(\widehat{\beta}^{\cdot})$  (the standard deviation of ${\beta}^{(m)}$ after burn-in and thinning), Bayes Factor estimate $\widehat{BF}_{01}$ for the Bayesian estimation, $p$-values for the frequentist estimation.}
	\label{tbl:beta_results}
\begin{tabular}{lccc} 
\toprule
Method & $\widehat{\beta}$ & $SE(\widehat{\beta})$ & $\widehat{BF}_{01}$/$p-value$ \\ \midrule
\multicolumn{4}{c}{ \textit{Sample 1}} \\
\midrule
BAY: $\aR$ random & 0.052 & 0.055 & 2.180 \\ 
RBE & 0.061 & 0.055 & 0.272 \\ 
OLS & 0.102 & 0.054 & 0.062 \\ 
\midrule
\multicolumn{4}{c}{ \textit{Sample 2}} \\
\midrule
BAY: $\aR$ random & 0.067 & 0.055 & 0.907 \\ 
RBE & 0.065 & 0.050 & 0.197 \\ 
OLS & 0.113 & 0.049 & 0.023 \\ 
\bottomrule
\end{tabular} 
\end{table}

\section{Conclusion} \label{sec:conclusion}

Return predictability is a frequently and also controversially debated question in the asset pricing literature
\citep[see, e.g.,][]{Campbell_book_2017,Golez_2018}. 
This article applies the restricted VAR model introduced in \citet{Cochrane_2008}, in which the log dividend-price ratio, $x_t$, follows a first-order autoregressive process, that is
$x_t = \alpha^x + \phi x_{t-1} + \varepsilon_{t}^x$,
and asset returns, $y_t$, are generated by means of $y_t = \alpha^y + \beta x_{t-1} + \varepsilon_{t}^y$. 
As pointed out already by \citet{Kendall1954}, the ordinary least squares estimator of $\phi$ is biased. This effect and the correlation of the noise terms result in biased ordinary least squares estimates of $\beta$ \citep[see][]{Stambaugh_1999}.

In this article, a Bayesian approach is developed to reduce these biases.  First, to improve the estimation of $\phi$, we impose a prior based on \citet{Berger_1994}. Second, to obtain a prior for the regression coefficient $\beta$ we apply a Beta-prior, with parameters denoted $\aR$ and $\bR$, on the coefficient of determination $R^2$ in the return equation. In addition to recent literature \citep[see][]{Cadonna_2020,Giannone_2021,Zhang_2020_r2_d2}, we impose a binary prior distribution on $\aR$. We demonstrate that this improves the performance of Bayesian testing. In particular, in a simulation study, we get more posterior mass on values $\aR$ with good false 
negative and false positive rates in the case of predictability and no predictability, respectively. 

We compare our Bayesian procedure to ordinary least squares estimation and the reduced-bias estimator proposed in \citet{Amihud_2004}.
The ordinary least squares estimator of $\beta$ is upward biased resulting in strong oversizing combined with high observed power. 
By choosing the parameters of our priors such that the false positive rate is close to the theoretical size of the frequentist test, the false negative rate obtained with our Bayesian approach is only slightly smaller than with ordinary least squares estimation. The Bayesian approach dominates the reduced-bias estimator in terms of both, observed size (false positive) and  power (false positive).
 
Finally, we apply our approach to annual CRSP value-weighted returns used in \citet{Cochrane_2008}, namely from January 1926 until December 2004 and returns from 1953 to 2021. For the first sample, the Bayesian test supports the hypothesis of no return predictability, while for the second dataset marginal evidence for predictability is observed.


\begin{doublespacing}
	\bibliographystyle{jf}
	\bibliography{library}
\end{doublespacing}

\appendices
\renewcommand{\thesection}{\Alph{section}}
\renewcommand{\thesubsection}{\Alph{section}.\Roman{subsection}}

\section{Priors} \label{app:priors}

\subsection[Explicit Prior]{Explicit Prior on $\Rsquare$}  \label{app:explicit_prior_r2}

We specify a direct prior on $\Rsquare$ similar to \citet{Giannone_2021} and \citet{Zhang_2020_r2_d2}. \citet{Giannone_2021} impose priors on standardized covariates, i.e. $\mathbb{E} \left( x_{jt} \right)=0$ and  $\mathbb{V} \left( x_{jt} \right) =1$ and obtain the following priors for the regression coefficients $\beta^\star_j$ in the standardized model:
\begin{eqnarray*}
	\beta^\star_j | \sigma_y^2  \sim \Normal{0, \sigma_y^2  \Sigmab },
\end{eqnarray*}
where $\Sigmab$ is fixed. For standardized covariates, \citet{Giannone_2021} define $\Rsquare$ unconditional on $\beta$:
\begin{align*}
	\Rsquare =  \frac{N \cdot \Sigmab}{N \cdot \Sigmab + N},
\end{align*}
where $N$ is the dimension of the regression parameter. For $N=1$, this reduces to
\begin{equation}\label{prior:explicit_r2}
	\Rsquare =  \frac{\Sigmab}{\Sigmab + 1}  \sim \Betadis{\aR,\bR}, \qquad \aR > 0, \bR > 0,
\end{equation}
where we impose a Beta prior on $\Rsquare$. The Beta distribution is the natural candidate for the prior on $\Rsquare$ since the support of the Beta distribution is $[0,1]$ and the Beta distribution is flexible enough to model different prior beliefs regarding $\Rsquare$. Using $\Sigmab= \Rsquare/(1-\Rsquare)$ and the relationship between the Beta distribution and the Beta prime distribution \citep[see, e.g.,][]{Johnson_book_1995}, we obtain the following hierarchical prior:
\begin{eqnarray*}
	\beta^\star | \sigma_y^2 , \Sigmab  \sim \Normal{0, \sigma_y^2  \Sigmab }, \qquad 	\Sigmab \sim  \Betapr{\aR ,\bR}.
	\end{eqnarray*}
Using \citet[Lemma~1]{Cadonna_2020}, it can be shown that this prior is equivalent to a triple gamma prior, also known as a normal-gamma-gamma prior \citep{Griffin_2017}, where we define $\Sigma_R=\frac{\bR}{\aR}
\Sigmab$: 
\begin{eqnarray*}
	\beta^\star | \sigma_y^2, \Sigma_R  \sim \Normal{0, \sigma_y^2 \frac{\aR}{\bR} \Sigma_R }, \qquad 	\Sigma_R \sim  \Fd{2\aR ,2\bR}.
\end{eqnarray*}
To apply this prior to our predictive system in Equation~\eqref{eq:model_1}, we first transform it such that $\Exp{x_{t}}=0$ and  $\Var{x_{t}}=1$ is standardized:
\begin{eqnarray*} 
	&&  y_t = (\alpY)^\star + \beta^\star \frac{(x_{t-1} - \alpX/(1 - \phi))}{\sqrt{\sigma^2_{xx}}} +  \epsilon _t^y,  \qquad \epsilon _t^y \sim \Normal{0, \sigma_y^2},
\end{eqnarray*}
where $\sigma^2_{xx} := \sigma_x^2/(1-\phi^2)$ is equal to the unconditional variance of the covariate $x_t$, 
where $\left( x_t \right)_{t \in \mathbb{Z}}$
following the AR(1)-process in Equation~\eqref{eq:model_1}. Matching priors yields the following prior for $\beta=\beta^\star/\sqrt{\sigma^2_{xx}}$:
\begin{eqnarray} \label{priorR2D2}
	\beta |  \phi,\sigma_x ^2, \sigma_y^2, \Sigmab \sim \Normal{0, \Sigmab \frac{\sigma_y^2 (1-\phi^2)}{\sigma_x ^2}}, \qquad  \Sigmab \sim  \Betapr{\aR ,\bR}.
\end{eqnarray}
This leads to an immediate extension of the prior of \citet{Wachter_2015} by allowing $\Sigmab$ (in \citet{Wachter_2015} notation this is $\sigma_{\eta}^2$) to be a random variable following the beta-prime distribution $\Sigmab \sim  \Betapr{\aR ,\bR}$ instead of assuming a fixed value.\footnote{Another important difference is that we do not condition on $\beta$ in the prior for $\Rsquare$.} Since we operate in model \eqref{eq:model_3}, we express  $\sigma_y^2/\sigma_x^2$ in terms of the parameters $(\psi, \sigmacon, \sigma_x^2)$  and define the prior variance for $\beta$ as:
\begin{align} \label{eqn:Sigma_0_beta}
    \Sigma^{\beta}_0 (\psi, \sigmacon, \sigma_x^2, \phi) = \Sigmab \frac{\sigma_y^2 (1 - \phi^2)}{\sigma_x^2} = 
     \Sigmab (1-\phi^2)  \left[\frac{\sigmacon}{\sigma_x ^2} + \psi^2\right]^{-1} .
\end{align}
The parameters $\aR$ and $\bR$ determine the prior shape of the prior distribution on $R^2$ and the induced prior on $\beta$ discussed in Section~\ref{sec:priors}. Figure~\ref{fig:prior_r2} presents the density plots for different values of $\aR$ and $\bR$.

\textsc{\begin{figure}[t]
		\includegraphics[width = 15cm]{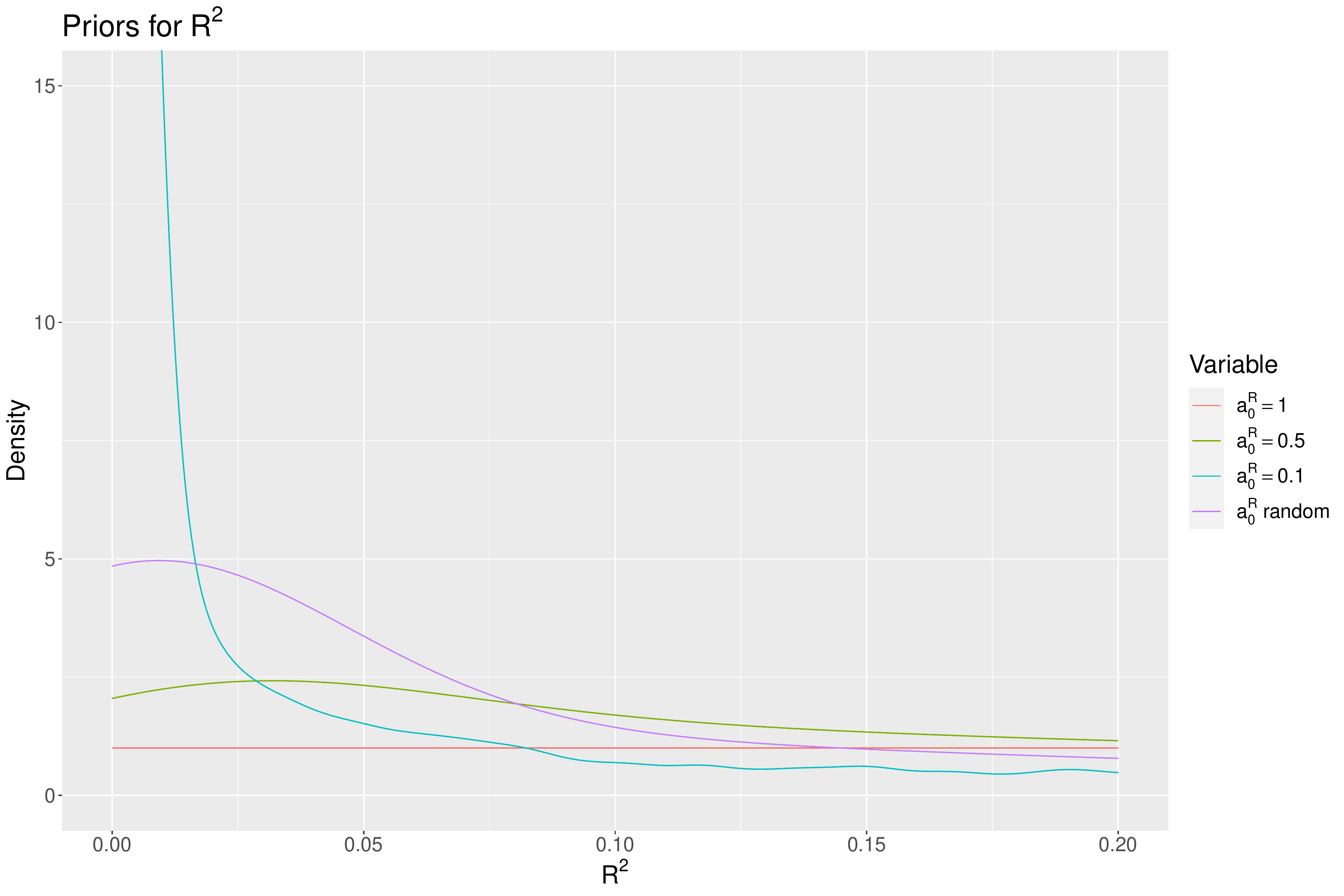}
		\centering
		\caption{Priors on $\Rsquare$. The prior on $\Rsquare$ is defined in Equation~\eqref{prior:explicit_r2} with different values for $\aR$ and $\bR$ being fixed to one. The first prior (red) is a flat prior with $\aR = 1$ and $\bR = 1$. The second prior (green) has $\aR = 0.5$ and $\bR = 1$. The third prior (cyan) has $\aR = 0.1$ and $\bR = 1$. The fourth prior (purple) has a hyperprior on $\aR$ with $\aR = \aRb = 0.5$ with probability $p_{\aR} = 0.5$ and $\aR = \aRu = 0.1$ with probability $1-p_{\aR} = 0.5$ discussed in Section~\ref{sec:priors}.}
		\label{fig:prior_r2}
\end{figure}}

\section{MCMC Details} \label{app:mcmc_details}

\subsection[Details for sampling.]{Details for Sampling $(\alpX, \phi)$} \label{app:mcmc_details_MH}

In this section, we discuss the sampling procedure for $\bm{\theta}_{1}|\mathbf{x},\tilde{\mathbf{u}},\tilde{\sigma}_x^2$, where $\bm{\theta}_{1} := (\alpX, \phi)$ and $\tilde{\sigma}_x^2$ is defined in (\ref{eq:step_1_3}). For a comprehensive review please refer to \citet{Kastner_2014}. We consider the \textit{centered parameterization}, namely Equation~\eqref{eq:model_1}. To sample $\bm{\theta}_{1}$, we apply an independence Metropolis-Hastings algorithm. To increase the MH acceptance rate, we employ an auxiliary regression model with the conjugate prior
\begin{align*}
	p_{aux}(\bm{\theta}_{1}|\tilde{\sigma}_x^2) \sim \mathcal{N}_2(\mathbf{b}_0, \tilde{\sigma}_x^2 \mathbf{B}_0)
\end{align*}
as auxiliary prior for $\bm{\theta}_{1}$ with
\begin{align*}
	\mathbf{B}_{0} &= \text{diag}(B_0^{11}, B_0^{22}), \ \text{ with } \ B_0^{11} = 10^{12}, B_0^{22} = 10^8,\\
	\mathbf{b}_{0} &= (0,0).
\end{align*}
This auxiliary prior allows us to derive the auxiliary conditional posterior distribution $\bm{\theta}_{1}|\mathbf{x},\tilde{\mathbf{u}},\tilde{\sigma}_x^2$,  given by:
\begin{equation} \label{eqn:appendix_full_conditional_theta_1}
	\bm{\theta}_{1}|\mathbf{x},\tilde{\mathbf{u}},\tilde{\sigma}_x^2 \sim \mathcal{N}_2(\mathbf{b}_T, \tilde{\sigma}_x^2\mathbf{B}_T),
\end{equation} 
with $\mathbf{B}_T = (\mathbf{X}^{'}\mathbf{X} + \mathbf{B}_{0}^{-1})^{-1}$ and $\mathbf{b}_T = \mathbf{B}_T[\mathbf{X}^{'}(\mathbf{x} - \tilde{\mathbf{u}}) + \mathbf{B}_{0}^{-1} \mathbf{b}_{0}]$, where $\mathbf{X}$ is a $T \times 2$ matrix with the $t$-th row given by $(1,\ x_{t-1})$ \footnote{Please note that $\mathbf{x} := (x_t)_{t=1,\dots,T}$}. 
Using the auxiliary posterior (\ref{eqn:appendix_full_conditional_theta_1}) as proposal in an independence MH algorithm,  a new value $\bm{\theta}_1 \new$ is proposed and accepted with probability  $\min(1,A)$, where
\begin{equation}\label{eqn:MH_R}
	A = \frac{p(x_0|\bm{\theta}_{1}\new, \sigma_x^2)}{p(x_0|\bm{\theta}_{1}, \sigma_x^2)} \times \frac{p\left(\beta| \Sigma_0^\beta ( \phi \new,  \cdot)\right)}{p\left(\beta| \Sigma_0^\beta ( \phi  , \cdot)\right)} \times \frac{p(\bm{\theta}_{1}\new)}{p(\bm{\theta}_{1}  )} \times \frac{p_{aux}(\bm{\theta}_{1}  )}{p_{aux}(\bm{\theta}_{1}\new)}.
\end{equation}
The first term in Equation~\eqref{eqn:MH_R} is the ratio of densities for the initial observation $x_0$ discussed in Equation~\eqref{eq:model_3}, the second term is the ratio of priors for $\beta$ since the prior $\beta| \phi \sim \Normal{0,\Sigma_0^\beta (\phi, \cdot)}$ depends on $\phi$ and serves as an additional likelihood term which has to be added to the acceptance rate, the third term is the ratio of priors that depends on the choice of the prior discussed in Section~\ref{sec:priors} and the fourth term in Equation~\eqref{eqn:MH_R} is the correction term that corresponds to the auxiliary priors employed to derive the 
auxiliary distribution $\bm{\theta}_{1}|\mathbf{x},\tilde{\mathbf{u}},\tilde{\sigma}_x^2$ in \eqref{eqn:appendix_full_conditional_theta_1}. The expression for $\log(A)$ becomes
\begin{align*}
	\log(A) &= \frac{1}{2}\log\left(\frac{1 - (\phi\new)^2}{1 - \phi ^2}\right) + \frac{1}{2\sigma_x^2}\left[(x_0 - \mu)^2 (1 - \phi  ^2) - (x_0 - \mu\new)^2 (1 - (\phi\new)^2)\right] \\
	&+ \log(\frac{1 - \phi  }{1 - \phi\new}) + \frac{1}{2\SalpX}\left[(\mu   - \mualpX)^2 - (\mu\new - \mualpX)^2 \right] + \frac{1}{2}\log\left(\frac{1 - \phi ^2}{1 - (\phi\new)^2}\right)\\
	&+ \frac{1}{2\tilde{\sigma}_x^2}[{\bm{\theta}_{1}\new}^{'} \mathbf{B}_0^{-1} \bm{\theta}_{1}\new - {\bm{\theta}_{1}  }^{'} \mathbf{B}_0^{-1} \bm{\theta}_{1}  ], \numberthis \label{eqn:MH_log_R}
\end{align*}
where we define $\mu := \alpX/(1 - \phi)$  and $\mu \new := (\alpX) \new/(1 - \phi\new )$ and $\log(A)$ stands for the natural logarithm of $A$.

\subsection[Sampling Details]{Sampling Details for $\Sigmab$} \label{app:mcmc_details_Sigmab}

Given $\Sigmab$, we first sample from the posterior distribution of $Z_{\beta}|\Sigmab$. Combing the prior distribution for $Z_{\beta}$ with the prior distribution for $\Sigmab$ conditional on $Z_{\beta}$,
\begin{align*}
	    p(Z_{\beta}) &\sim \mathcal{G}(a_R, 1) \propto Z_{\beta}^{a_R - 1} \exp(-Z_{\beta}),\\
	    p(\Sigmab |Z_{\beta}) &\sim \mathcal{IG}(b_R, Z_{\beta}) \propto \left(\frac{1}{\Sigmab }\right)^{b_R + 1} \times \exp\left(-\frac{Z_{\beta}}{\Sigmab }\right) \times Z_{\beta}^{b_R},
\end{align*}
the posterior distribution of $Z_{\beta}| \Sigmab$ is given by
\begin{align*}
  p(Z_\beta|\Sigmab ) &\propto p(\Sigmab  | Z_\beta) p(Z_\beta) \propto
    Z_\beta ^{\bR}  \exp \left(-\frac{Z_\beta }{\Sigmab } \right) Z_\beta^{\aR-1} \exp (-Z_\beta) \\
    &\propto  p(Z_\beta|\Sigmab) \propto Z_\beta ^{\bR + \aR - 1}  \exp \left( -Z_\beta \left(1 + \frac{1}{\Sigmab} \right) \right) \\
    &\sim \mathcal{G} \left( \aR + \bR, 1 + \frac{1}{\Sigmab} \right) \; .
\end{align*}
Next, we sample from the posterior distribution of $\Sigmab$, combining the prior distribution for $\beta$ with
{the} prior distribution of $\Sigmab$ conditional on $Z_{\beta}$:
\begin{align*}
	    p(\beta| \Sigmab ,\phi,\sigma_x ^2, \sigmacon, \psi) &\sim \mathcal{N}\left(0, \Sigmab
     (1-\phi^2)  \left[\frac{\sigmacon}{\sigma_x ^2} + \psi^2\right]^{-1}  \right) , \\
	    p(\Sigmab | Z_\beta) &\sim \mathcal{IG}(b_R, Z_{\beta}) \propto \left( \frac{1}{\Sigmab } \right)^{b_R + 1} \times \exp \left(-\frac{Z_{\beta}}{\Sigmab } \right) \ .
\end{align*}
Hence,  the posterior for $\Sigmab$ is given
\begin{align*}
p(\Sigmab | \cdot ) &\propto p(\beta| \Sigmab ,\phi,\sigma_x ^2, \sigmacon, \psi) \times p(\Sigmab | Z_\beta) \\
&\propto
\left(\frac{1}{\Sigmab }\right)^{\bR + \frac{1}{2} + 1} \times \exp \left( -\frac{1}{\Sigmab }\left[\frac{\beta^2}{2(1 - \phi^2)}\left(\frac{\sigmacon}{\sigma_x ^2} + \psi^2 \right)^{-1} + Z_{\beta}\right] \right),
\end{align*}
which results in 
\begin{align*} 
    p(\Sigmab| \cdot)     \sim \Gammainv{\bR + \frac{1}{2}, Z_\beta + S_\beta }  \quad \text{ and } \quad S_\beta=\frac{\beta^2}{2 (1-\phi^2)}  \left( \frac{\sigmacon}{\sigma_x ^2} + \psi^2\right)^{-1}.
\end{align*}
%

\subsection[Sampling Details]{Sampling of $\aR$} \label{app:mcmc_details_aR}

In this section, we discuss the sampling procedure for $p(\aR|\Sigma_{\beta})$, where we apply a MH step with acceptance rate $\min\{1,A\}$, where
\begin{align*}
    A = \frac{p\left(\aRnew|\Sigma_{\beta}\right)}{p\left(\aRold|\Sigma_{\beta}\right)} \times \frac{q\left(\aRold|\aRnew\right)}{q\left(\aRnew|\aRold\right)},
\end{align*}
where the first term is the posterior ratio and the last term is the ratio of the proposal densities. We choose a symmetric proposal density $q\left(\aRold|\aRnew\right) = q\left(\aRnew|\aRold\right)$, where the proposal ratio cancels. For the first term we can write 
\begin{align*}
    \frac{p\left(\aRnew|\Sigma_{\beta} \right)}{p\left(\aRold|\Sigma_{\beta}\right)} = \frac{p\left(\Sigma_{\beta}|\aRnew\right) p\left(\aRnew\right)}{p\left(\Sigma_{\beta}|\aRold\right) p\left(\aRold\right)},
\end{align*}
where $p\left(\Sigma_{\beta}|\aR \right)$ is the density of a Beta prime distribution:
\begin{align*}
p\left(\Sigma_{\beta}|\aR\right)=
\frac{1}{B(\aR,\bR)} (\Sigma_{\beta}) ^{\aR-1} \left(1+ \Sigma_{\beta} \right)^{-(\aR+\bR)} .
\end{align*}
This follows from the fact that only the prior of $\Sigma_{\beta}$ depends on $\aR$, while the remaining parameters are independent of $\aR$, given $\Sigma_{\beta}$. 
Given a binary prior distribution for $\aR$ with two values with equal probability, we obtain the expression
\begin{align} \label{ASylvia}
   & \log(A) =  
   \left(\aRnew - \aRold \right)\left[\log(\Sigma_{\beta})  -
   \log(1 + \Sigma_{\beta})\right]\\
   & +
   \log \Gamma\left(\aRold \right)- 
   \log \Gamma\left(\aRold +\bR\right)
   + \log \Gamma\left(\aRnew+\bR\right)
 -  \log \Gamma\left(\aRnew\right)
   \nonumber,
\end{align}
and $\Gamma \left( z \right)$ denotes the Euler's Gamma function.

\subsection{Convergence Diagnostics}
\label{app:mcmc_details_convergence}

To ensure satisfactory performance of the sampler we perform a convergence analysis for the $M$ thinned posterior draws 
$\xi^{(m)}_i$, $m=1,\dots,M$, for each 
data set $\mathbb{D}_{T,i}$, $i = 1,\ldots, n $, on a separate basis.
By following \citet{Gelman_book_1995}[Chapter~11.5]
the {\em Effective sample size} (ESS) is obtained by means of
\begin{align*}
	{M}^{eff}_i := \frac{M}{1 + 2 \sum_{\ell =1}^{\infty} {\rho}_{i,\ell}} \ ,
\end{align*}
where is ${\rho}_{i,\ell}$ is the autocorrelation of $\xi^{(m)}_i$ at lag $\ell$. To estimate the ``long run correlation term`` in the denominator we use the \texttt{coda} package\footnote{See the package website: \href{https://cran.r-project.org/web/packages/coda/coda.pdf}{https://cran.r-project.org/web/packages/coda/coda.pdf}} in \texttt{R}, which relies on estimating the spectral density at frequency zero. 
This results in the estimates $\widehat{M}^{eff}_i$, $i=1,\dots,n$.
We exclude those $n-n_d$ data sets $\mathbb{D}_{T,i}$ where {$\widehat{M}_i^{eff}$} is less than a third of the actual sample size after burn-in and thinning, denoted by $M$ in the main text. 
In addition, we analyze visually the trace plots for $\xi_i^{(m)}$ to confirm that mixing is satisfactory and we have reached the stationary distribution of the Markov chain considered.

\section{Bayesian Testing} \label{app:simulation}

\subsection{Computation of Bayes Factors via the Savage-Dickey Density Ratio} \label{app:simulation_details_BF}

Since the prior density $p(\beta)$ defined in Section~\ref{sec:priors} has no closed form we use the hierarchical representation,
\begin{align*}
	p(\beta|\bm{\psi}) \sim \mathcal{N}\left(0, g(\bm{\psi})\right), \text{ where } \bm{\psi} \sim p(\bm{\psi}),
\end{align*}
and $g(\bm{\psi})$ is a non-linear function of the model parameters $\bm{\psi}$. The prior $p(\beta)$ can then be approximated by
\begin{align*}
	p(\beta) = \int p(\beta|\bm{\psi})p(\bm{\psi})d\bm{\psi} = \lim_{M \to \infty}\frac{1}{M}\sum_{m=1}^M p_{\mathcal{N}}
 \left(\beta; 0, g(\bm{\psi}^{(m)})\right),
\end{align*}
where $p_{\mathcal{N}}(\beta; 0, S)$ is the pdf of a Gaussian distribution with mean 0 and variance $S$ and $\bm{\psi}^{(m)} \sim p(\bm{\psi})$ are independent draws from $p(\bm{\psi})$. To approximate the prior ordinate $p(\beta = 0)$, we evaluate the densities at $p_N\left(0; 0, g(\bm{\psi}^{(m)})\right)$ and obtain
\begin{align}
\label{eq:BF:pibeta0}
	p(\beta = 0) = \lim_{M \to \infty} \frac{1}{M \sqrt{2\pi}}\sum_{m=1}^M \frac{1}{\sqrt{g(\bm{\psi}^{(m)})}}.
\end{align}
The posterior density of $p(\beta|\mathbb{D}_{T})$ has the following representation
\begin{align*}
	p(\beta|\mathbb{D}_{T}) = \int p(\beta,\bm{\psi}|\mathbb{D}_{T})d\bm{\psi} = \int p_N(\beta|\bm{\psi},\mathbb{D}_{T}) p(\bm{\psi}|\mathbb{D}_{T})d\bm{\psi},
\end{align*}
where $\bm{\psi}$ are all parameters that appear in the conditioning argument in Step~1 of the MCMC sampler described in Algorithm~\ref{Algo1}:
\begin{align*}
	\beta|\bm{\psi},\mathbb{D}_{T} \sim \mathcal{N}\left(\beta; b_T(\bm{\psi}), B_T(\bm{\psi})\right),
\end{align*}
where $b_T$ is the second element of $\mum_{T}$ and $B_T$ is the second diagonal element of $\Sigmam_{T}$ defined in Equation~(\ref{eq:mcmc_full_posterior_beta}). Therefore $p(\beta|\mathbb{D}_{T})$ can be approximated by
\begin{align*}
	p(\beta|\mathbb{D}_{T}) = \lim_{M \to \infty} \frac{1}{M}\sum_{m=1}^M p_N\left(\beta; b_T^{(m)}, B_T^{(m)}\right),
\end{align*}
where $b_T^{(m)}$ and $B_T^{(m)}$ are the conditional posterior moments, when sampling $\beta^{(m)}$ at the $m$-th iteration of the sampler. To approximate the posterior ordinate $p(\beta = 0|\mathbb{D}_{T})$,  we evaluate the densities at $p_{\mathcal{N}}(0; b_T^{(m)}, B_T^{(m)})$ and obtain
\begin{align}
	\label{eq:BF:pibeta0y}
 p(\beta = 0|\mathbb{D}_{T}) = \lim_{M \to \infty} \frac{1}{M \sqrt{2\pi}} \sum_{m=1}^M \frac{1}{\sqrt{B_T^{(m)}}} \exp\left[-\frac{(b_T^{(m)})^2}{2B_T^{(m)}}\right]
\end{align}
By combining 
(\ref{eq:BF:pibeta0}) and 
(\ref{eq:BF:pibeta0y}), we could estimate $BF_{01}$ in (\ref{eq:savage_density_ratio}) by the finite sample estimators of prior and the posterior ordinate at $0$:
\begin{align*}
\widehat{BF}_{01} = 
\frac{\frac{1}{M}
\sum_{m=1}^M \frac{1}{\sqrt{B_T^{(m)}}} \exp
\left[ -\frac{ \left(b_T^{(m)} \right)^2}{2B_T^{(m)}} \right]
}
{ \frac{1}{M}
\sum_{m=1}^M \frac{1}{\sqrt{g\left( \bm{\psi}^{(m)} \right) } } } \ .
\end{align*}
However, given that the distribution might be right-skewed we use the median estimator instead of the mean [the posterior median minimizes the error under the $L_1$ loss function]. This results in the following estimator of $BF_{01}$:
\begin{align} \label{eq:BF01}
\widehat{BF}_{01} = 
\frac{
Q_{0.5,M} \left( \frac{1}{\sqrt{B_T^{(m)}}} \exp
\left[ -\frac{ \left(b_T^{(m)} \right)^2}{2B_T^{(m)}} \right] \right)}
{Q_{0.5,M} \left(\frac{1}{\sqrt{g\left( \bm{\psi}^{(m)} \right) } } \right)} \ ,
\end{align}
where $Q_{0.5,M} (\xi^{(m)})$ denotes the median of the $M$ draws of $\xi^{(m)}$.

\subsection{An Example on Bayes Factors} \label{app:simulation_details_BF_example}

We illustrate the idea of the ${\widehat{FP}}$ and ${\widehat{FN}}$ measures that are based on the Bayes factor in Figure~\ref{fig:BF_example}. The blue line is the hypothetical distribution of the Bayes Factor under {\em DGP 1} and the red line is the hypothetical distribution of the BF under {\em DGP 0}. Given the value $BF_{01}$ and the threshold $C$ we have a decision rule whether we have predictability or not. If {$BF_{01} < (\geq) \ K=1$} we decide that we have (no) predictability because the posterior distribution is smaller (larger) than the prior distribution in Equation~\eqref{eq:savage_density_ratio}. The filled blue and red areas indicate the ${\widehat{FN}}$ and ${\widehat{FP}}$ rates respectively. For example, the integral of the red area shows the probability of observing $BF_{01} < K$ and deciding that we have predictability given that we have {\em DGP 0} (False Positive rate). Alternatively, the integral of the blue area shows the probability of observing $BF_{01} > K$ and deciding that we have no predictability given that we have {\em DGP 1} (False Negative rate). These two measures are especially important since our ultimate goal is to test whether the dividend-price ratio predicts future returns (or not). This means that for real data we might not be interested in the value of the coefficient in the predictive regression per se, but rather if it is statistically significant or not. 

\textsc{\begin{figure}[t]
		\includegraphics[width = 15cm]{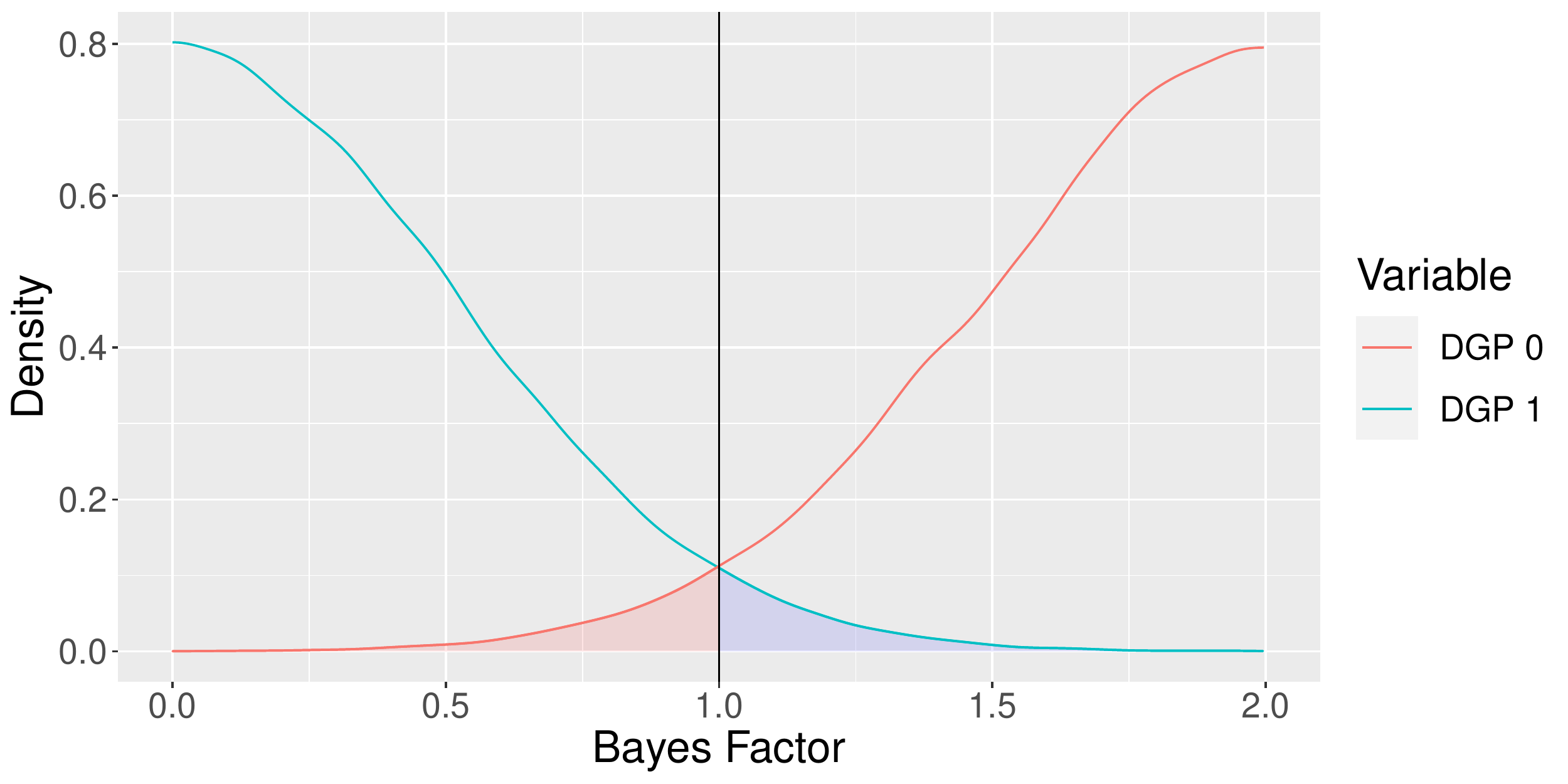}
		\centering
		\caption{Hypothetical distribution of the Bayes Factors under {\em DGP 0} (red line) and {\em DGP 1} (blue line). The vertical line indicates the testing threshold ${K} = 1$. The red area represents the size of the test and the blue area represents the power of the test.}
		\label{fig:BF_example}
\end{figure}}

\subsection{Simulation Results on Bayes Factors} \label{app:simulation_BF_result}

We analyze the result for ${\widehat{FP}}$ and ${\widehat{FN}}$ in more details in Figure~\ref{fig:BF_bayes} that shows the sampling distribution of the Bayes Factor for {\em DGP 0} and {\em DGP 1} similar to the example in Figure~\eqref{fig:BF_example}. The vertical line (equals 1) indicates the threshold $K$ above which we conclude that there is no predictability. In Figure~\ref{fig:BF_bayes} we analyze the sampling distribution of the $t$-values for {\em DGP 0} and {\em DGP 1} for OLS and RBE. The vertical lines (equal to $-1.96$ and $1.96$) indicate the standard thresholds when we reject/non-reject the null hypothesis of no predictability 
{at the 5\% significance level}. The filled blue area indicates the ${\widehat{FN}}$ rate that shows the situation when we conclude that there is no predictability whereas we do have one. The red filled area represents the ${\widehat{FP}}$, meaning that we conclude that there is predictability, whereas we do not have one.  For the OLS we have violations for the right tail of the sampling distribution under {\em DGP 0}, meaning that we detect predictability for $\beta=0$. For RBE we decrease the bias by shifting the sampling distribution of $\beta$ close to zero. This decreases most of the $t$-values, but $t$-values smaller than $-1.96$ contribute to the ${\widehat{FP}}$ rate. {Under Bayesian testing, we observe that the sampling distribution of the Bayes Factor for \textit{DGP 0} is rather flat, whereas for \textit{DGP 1} it concentrates at values close to zero.}

\textsc{\begin{figure}[t!]
		\includegraphics[width = 15cm]{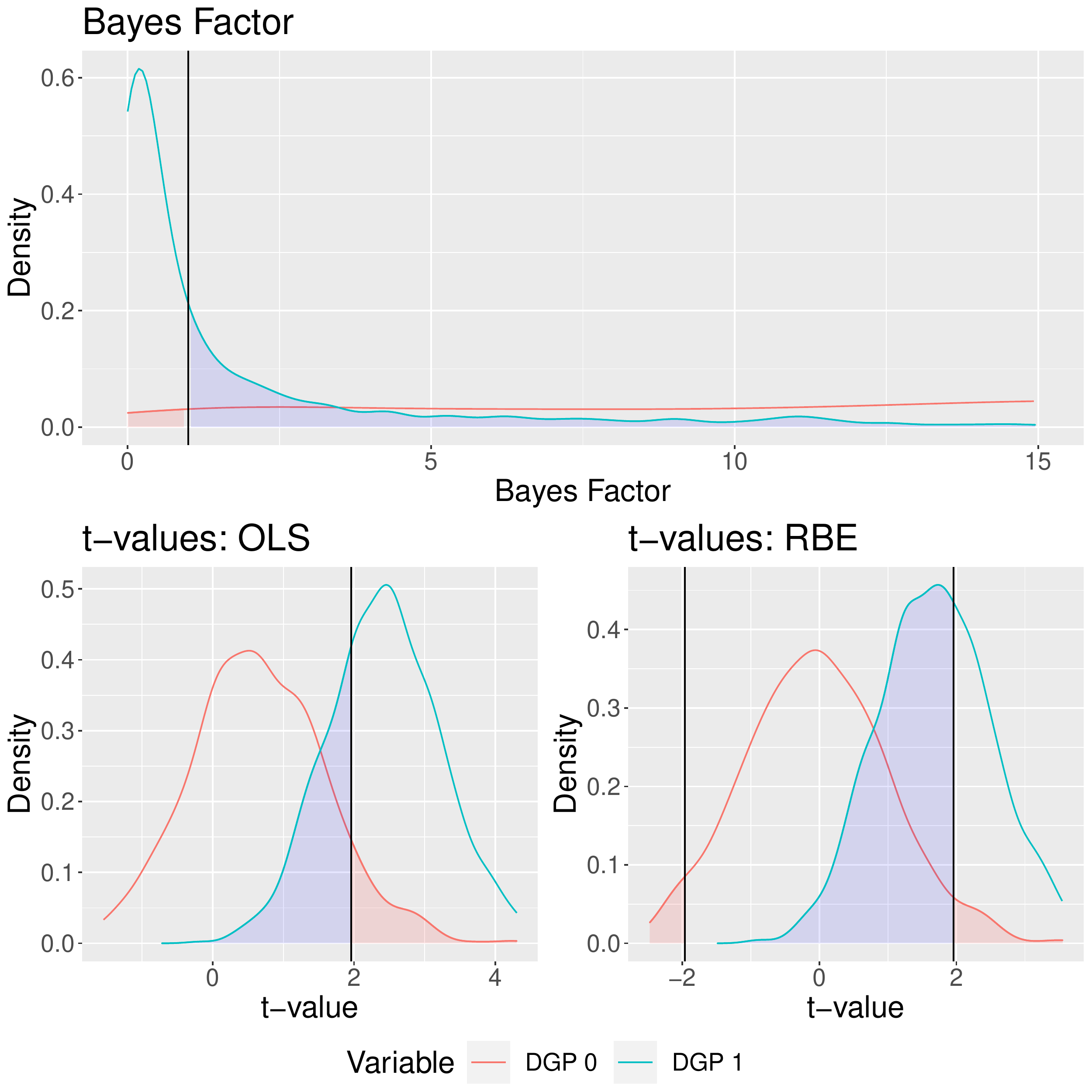}
		\centering
		\caption{Sampling distribution of the Bayes Factor for the parameter ${\beta}$ for the Bayesian estimator (upper panel) and sampling distribution for the $t$-values for the frequentist estimators OLS (lower left panel) and RBE (lower right panel). The left panel is the result for {\em DGP 0} and the right panel is the result for {\em DGP 1}.}
		\label{fig:BF_bayes}
\end{figure}}

\section[Robustness Checks]{Robustness Checks regarding $\beta$ and $\aR$} 
\label{app:sim_res_diff_beta}

We analyze additional cases where $\beta \in \{0, 0.025, 0.05, 0.075, 0.1, 0.2\}$ and denote the corresponding simulation settings as $DGP_i$ for $i \in \{0, \ldots, 5\}$ respectively. Additionally, we analyze the case when the prior parameter $\aR$ is fixed and equals either $0.1$ or $0.5$ and $\bR$ equals 1. We compare the performance of these two methods with the results presented in Table~ \ref{tbl:beta_sim_data_rmse}. We investigate how the performance of the methods changes if we change the level of predictability. We expect that for smaller values of $\beta$ it would be harder to reject the null and for larger values of $\beta$ easier to reject the null. The results presented in Table~\ref{tbl_app:beta_full_sim_data_rmse}
{confirm these expectations. The estimated false negative rates (observed power) are monotonically increasing in $\beta$. 
The observed power is high for OLS.
For larger values of $\beta$ the difference between the Bayesian estimator and OLS decreases and for $\beta = 0.2$ we have even better-observed power for the Bayesian estimator compared to OLS. } In terms of MAE and RMSE, the results are stable for OLS and RBE, while deteriorating for the Bayesian estimator. This follows from the choice of a prior that puts a lot of mass on small values of $\beta$, affecting the posterior.

\begin{table}[H] \centering 
	\caption{Measures of the estimation quality for $\widehat{\beta}$ (multiplied by 100): Bias, Standard deviation, MAE, RMSE, estimated False Positive ($\widehat{FP}$) and estimated False Negative rates ($\widehat{FN}$). Three estimation methods: OLS, RBE and BAY. We have 6 {\em DGPs}  where $\beta \in \{0, 0.025, 0.05, 0.075, 0.1, 0.2\}$ and $\bR = 1$.} 
	\label{tbl_app:beta_full_sim_data_rmse} 
	\begin{tabular}{lccccc} 
 \toprule
		Method & $B(\widehat{\beta}, \beta$) & $\sigma(\widehat{\beta}$) & MAE($\widehat{\beta}, \beta$) & RMSE($\widehat{\beta}, \beta$) & - \\ \midrule
		\multicolumn{5}{c}{$\beta = 0$} & $\widehat{FP}$ \\
		\midrule
OLS & 4.33 & 6.56 & 5.77 & 7.86 & 8.14 \\ 
RBE & 0.47 & 6.69 & 5.1 & 6.71 & 7.21 \\ 
BAY: $\aR$ random & 1.94 & 3.93 & 2.55 & 4.38 & 6.14 \\ 
BAY: $\aR = 0.1$ & 1.32 & 2.69 & 1.62 & 2.99 & 7.36 \\ 
BAY: $\aR = 0.5$ & 3.32 & 5.09 & 4.17 & 6.07 & 4.52 \\ 
		\midrule
		\multicolumn{5}{c}{$\beta = 0.025$} & $\widehat{FN}$\\
		\midrule
OLS & 4.09 & 6.39 & 5.5 & 7.58 & 86.1 \\ 
RBE & 0.23 & 6.52 & 4.94 & 6.52 & 94.4 \\ 
BAY: $\aR$ random & 0.8 & 4.51 & 2.83 & 4.58 & 88.79 \\
BAY: $\aR = 0.1$ & -0.25 & 3.66 & 2.39 & 3.67 & 86.75 \\ 
BAY: $\aR = 0.5$ & 2.52 & 5.28 & 4.08 & 5.85 & 91.88 \\
        \midrule
		\multicolumn{5}{c}{$\beta = 0.05$} & $\widehat{FN}$\\
		\midrule 
OLS & 4.49 & 6.68 & 5.9 & 8.05 & 70.17 \\ 
RBE & 0.61 & 6.81 & 5.2 & 6.84 & 88.76 \\ 
BAY: $\aR$ random & 0.2 & 5.41 & 3.89 & 5.41 & 77.03 \\  
BAY: $\aR = 0.1$ & -1.26 & 4.9 & 3.82 & 5.06 & 73.4 \\ 
BAY: $\aR = 0.5$ & 1.8 & 5.9 & 4.26 & 6.17 & 85.69 \\
        \midrule
		\multicolumn{5}{c}{$\beta = 0.075$} & $\widehat{FN}$\\
		\midrule
OLS & 4.64 & 6.38 & 5.92 & 7.88 & 48.4 \\ 
RBE & 0.75 & 6.5 & 4.91 & 6.54 & 80 \\ 
BAY: $\aR$ random & 0.09 & 5.96 & 4.5 & 5.95 & 58.51 \\ 
BAY: $\aR = 0.1$ & -2.38 & 5.06 & 4.83 & 5.59 & 57.75 \\ 
BAY: $\aR = 0.5$ & 1.95 & 6.26 & 4.73 & 6.56 & 70.36 \\ 
        \midrule
		\multicolumn{5}{c}{$\beta = 0.1$} & $\widehat{FN}$\\
		\midrule
OLS & 4.46 & 6.67 & 5.89 & 8.02 & 28.3 \\ 
RBE & 0.6 & 6.8 & 5.2 & 6.83 & 63.27 \\ 
BAY: $\aR$ random & 0.27 & 6.83 & 5.16 & 6.84 & 36.19 \\
BAY: $\aR = 0.1$ & -2.11 & 6.67 & 5.78 & 6.99 & 34.91 \\ 
BAY: $\aR = 0.5$ & 1.58 & 6.2 & 4.82 & 6.39 & 52.98 \\ 
        \midrule 
		\multicolumn{5}{c}{$\beta = 0.2$} & $\widehat{FN}$ \\
		\midrule
OLS & 4.34 & 6.6 & 5.82 & 7.9 & 0.39 \\ 
RBE & 0.47 & 6.73 & 5.18 & 6.74 & 3.27 \\ 
BAY: $\aR$ random & 1.86 & 6.65 & 5.14 & 6.91 & 0.3 \\ 
BAY: $\aR = 0.1$ & 1.22 & 6.6 & 5.07 & 6.71 & 0.53 \\ 
BAY: $\aR = 0.5$ & 2.72 & 6.45 & 5.2 & 7 & 1.3 \\
\bottomrule
	\end{tabular} 
\end{table}

\begin{table}[H] \centering 
	\caption{Measures of the estimation quality for $\widehat{\beta}$ (multiplied by 100): Bias, Standard deviation, MAE, RMSE, estimated False Positive ($\widehat{FP}$) and estimated False Negative rates ($\widehat{FN}$). Three estimation methods: OLS, RBE and BAY. We have 6 {\em DGPs}  where $\beta \in \{0, 0.025, 0.05, 0.075, 0.1, 0.2\}$ and $\bR = 0.5$.} 
	\label{tbl_app:beta_full_sim_data_rmse_v2} 
	\begin{tabular}{lccccc}
 \toprule
		Method & $B(\widehat{\beta}, \beta$) & $\sigma(\widehat{\beta}$) & MAE($\widehat{\beta}, \beta$) & RMSE($\widehat{\beta}, \beta$) & - \\ \midrule
		\multicolumn{5}{c}{$\beta = 0$} & $\widehat{FP}$\\
		\midrule
OLS & 4.33 & 6.56 & 5.77 & 7.86 & 8.14 \\ 
RBE & 0.47 & 6.69 & 5.1 & 6.71 & 7.21 \\ 
BAY: $\aR$ random & 2.03 & 4.29 & 2.65 & 4.74 & 4.3 \\ 
BAY: $\aR = 0.1$ & 1.59 & 3.46 & 1.91 & 3.81 & 7.5 \\ 
BAY: $\aR = 0.5$ & 3.1 & 4.87 & 4.04 & 5.77 & 1.91 \\ 
		\midrule
		\multicolumn{5}{c}{$\beta = 0.025$} & $\widehat{FN}$\\
	    \midrule
OLS & 4.09 & 6.39 & 5.5 & 7.58 & 86.1 \\ 
RBE & 0.23 & 6.52 & 4.94 & 6.52 & 94.4 \\ 
BAY: $\aR$ random & 1.09 & 4.82 & 3.08 & 4.94 & 91.48 \\ 
BAY: $\aR = 0.1$ & 0.1 & 4.13 & 2.52 & 4.13 & 87.7 \\ 
BAY: $\aR = 0.5$ & 2.68 & 5.48 & 4.09 & 6.1 & 94.78 \\ 
        \midrule
		\multicolumn{5}{c}{$\beta = 0.05$} & $\widehat{FN}$\\
		\midrule 
OLS & 4.49 & 6.68 & 5.9 & 8.05 & 70.17 \\ 
RBE & 0.61 & 6.81 & 5.2 & 6.84 & 88.76 \\ 
BAY: $\aR$ random & 0.46 & 5.68 & 4.12 & 5.69 & 80.72 \\ 
BAY: $\aR = 0.1$ & -0.85 & 5.06 & 3.82 & 5.12 & 73.42 \\ 
BAY: $\aR = 0.5$ & 1.91 & 5.75 & 4.35 & 6.05 & 91.13 \\
        \midrule
		\multicolumn{5}{c}{$\beta = 0.075$} & $\widehat{FN}$\\
		\midrule 
OLS & 4.64 & 6.38 & 5.92 & 7.88 & 48.4 \\ 
RBE & 0.75 & 6.5 & 4.91 & 6.54 & 80 \\ 
BAY: $\aR$ random & 0.42 & 6.26 & 4.74 & 6.27 & 66.43 \\ 
BAY: $\aR = 0.1$ & -1.6 & 5.72 & 4.87 & 5.94 & 57.49 \\ 
BAY: $\aR = 0.5$ & 2.04 & 6.59 & 4.89 & 6.89 & 79.25 \\ 
        \midrule 
		\multicolumn{5}{c}{$\beta = 0.1$} & $\widehat{FN}$\\
		\midrule
OLS & 4.46 & 6.67 & 5.89 & 8.02 & 28.3 \\ 
RBE & 0.6 & 6.8 & 5.2 & 6.83 & 63.27 \\ 
BAY: $\aR$ random & 1.04 & 7.06 & 5.46 & 7.13 & 43.22 \\  
BAY: $\aR = 0.1$ & -0.91 & 6.79 & 5.64 & 6.85 & 31.99 \\ 
BAY: $\aR = 0.5$ & 1.95 & 6.14 & 4.65 & 6.44 & 63.76 \\ 
        \midrule
		\multicolumn{5}{c}{$\beta = 0.2$} & $\widehat{FN}$\\
		\midrule
OLS & 4.34 & 6.6 & 5.82 & 7.9 & 0.39 \\ 
RBE & 0.47 & 6.73 & 5.18 & 6.74 & 3.27 \\ 
BAY: $\aR$ random & 2.17 & 6.7 & 5.28 & 7.03 & 1.7 \\ 
BAY: $\aR = 0.1$ & 1.67 & 6.6 & 5.18 & 6.81 & 0.4 \\  
BAY: $\aR = 0.5$ & 2.83 & 6.25 & 5.03 & 6.86 & 2.8 \\
\bottomrule
	\end{tabular} 
\end{table}

\subsection[Different Values]{Different Values for $\beta$ and $T$} \label{app:sim_res_diff_beta_and_T_len}

In Figure~\ref{fig:beta_avg_posterior_v2} we present the estimated marginal posterior distributions $\widehat{p}\left({\beta}|\mathbb{D}_{T,i} \right)$ by the kernel density estimator using the thinned draws $\beta^{(m)}_i$, $m=1,\dots,2\,000$ for $i=1,\dots,n_d$ for three \textit{DGP}, where we increase the sample size $T$ and/or the value of $\beta$. The results indicate that the effect of the prior decreases as more observations or a larger level of predictability is observed.

\textsc{\begin{figure}[t!]
		\includegraphics[width = 10cm]{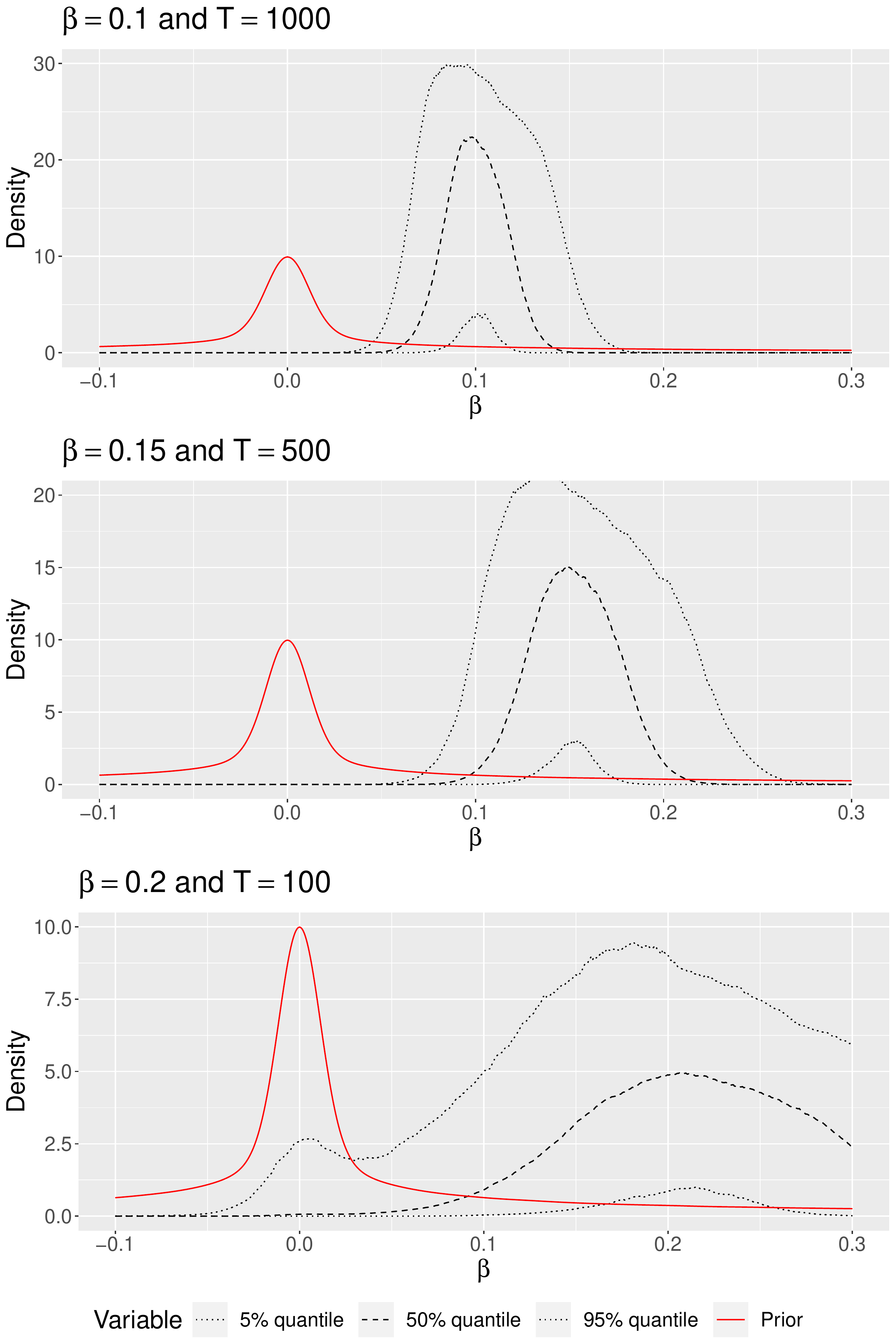}
		\centering
		\caption{5/50/95\% quantiles (dashed lines) of the estimated marginal posterior distributions $\widehat{p}\left({\beta}|\mathbb{D}_{T,i} \right)$ by the kernel density estimator implemented in {\tt R}, using default bandwidth choice using the thinned draws $\beta^{(m)}_i$, $m=1,\dots,2\,000$ for $i=1,\dots,n_d$. The top panel is the result for $\beta = 0.1$ and $T = 1\,000$. The middle panel is the result for $\beta = 0.15$ and $T = 5\,000$. The bottom panel is the result for $\beta = 0.2$ and $T = 100$.
}
\label{fig:beta_avg_posterior_v2}
\end{figure}}

\section{Financial Data} \label{app:financial_data}

\subsection{Testing Stationarity} \label{app:ar_1.phi}

\textsc{\begin{figure}[t!]
		\includegraphics[width = 12cm]{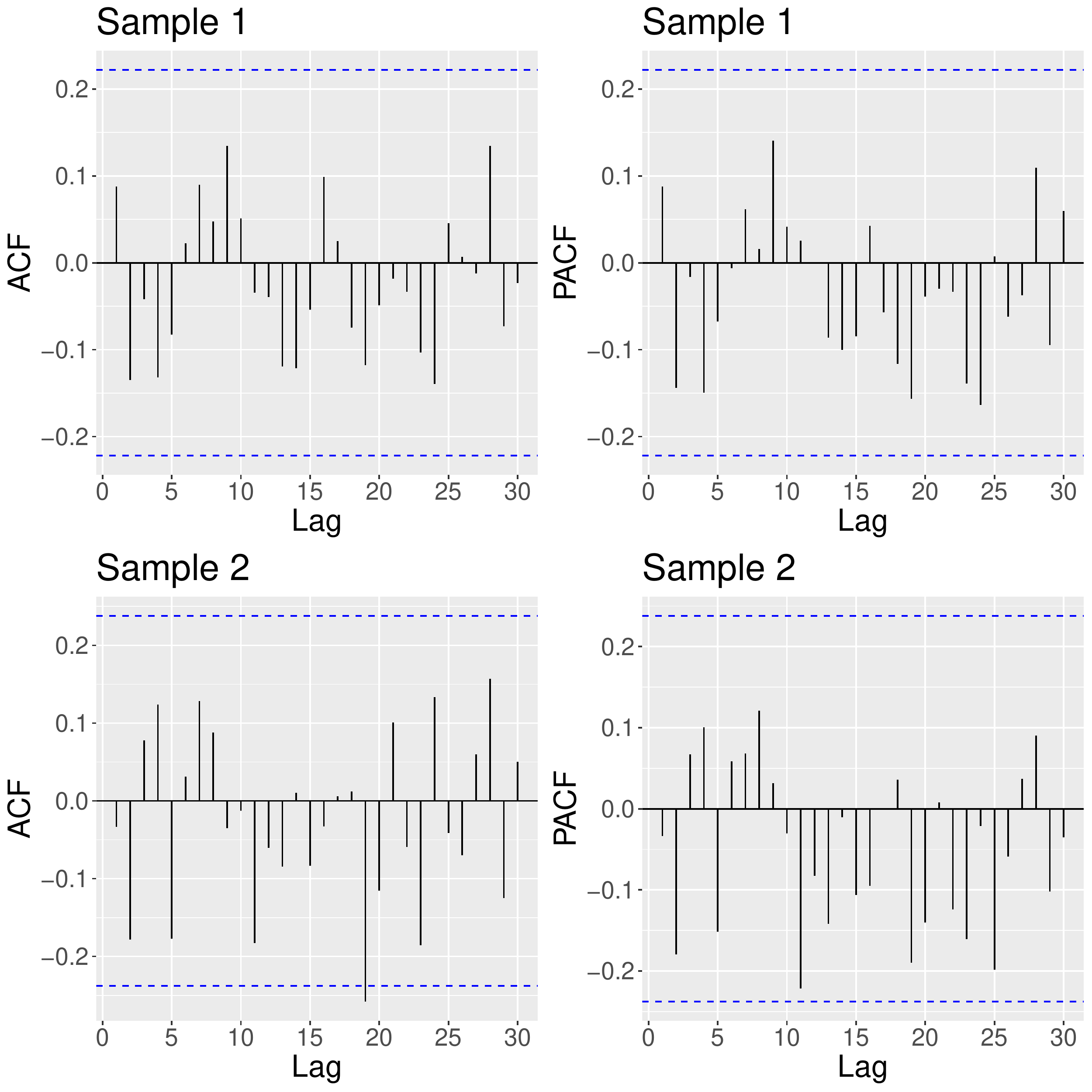}
		\centering
		\caption{ACF and PACF plots for log returns $y_t$. The upper panel is the result for {\em Sample 1} and the lower panel is the result for {\em Sample 2}. The left panel is the ACF plot and the right panel is the PACF plot. {The dotted lines show 95\% confidence bands.}
  }
\label{fig:acf_pacf_ret}
\end{figure}}

\noindent This section briefly investigates whether the data $\mathbb{D}_T$ is stationary. For simulated data, we considered a solution of the linear stochastic difference equation (\ref{eq:model_1})
on $\mathbb{Z}$, where $|\phi| < 1$ results in a stationary stochastic process $\left(x_t, y_t \right)^{\prime}_{t \in \mathbb{Z}}$. [Since the noise terms $\epsilon_t$ are $iid.$ normally distributed we can establish weak and strong stationarity.] For our empirical data sets considered in 
Section~\ref{sec:financial} we investigate this question by 
visual inspection of the autocorrelation function (ACF) and the partial autocorrelation function (PACF), and running augmented Dickey-Fuller (ADF) tests 
\citep[][]{dickey:1979}
as well as the KPSS test
\citep[][]{kpss:1992}. 
Here the EVIEWS~13 package was used.

{\it Returns $y_t$:} For the empirical returns used in this article, all autocorrelations as well as partial autocorrelations for lag orders $\leq 30$ are insignificant; see Figure~\ref{fig:acf_pacf_ret}. For the augmented Dickey-Full test we reject the null hypothesis of a unit root at a 1\% significance level, while for the KPSS test the null hypothesis that the time-series is integrated of order zero is not rejected at  a significance level of 10\%. These results are robust in the number of lags used for the augmented Dickey-Fuller test and in the bandwidth used to obtain an estimate of the long-run variance in the case of the KPSS test.

{\it Log dividend-price ratio $x_t$:} 
The autocorrelations and the partial autocorrelations have the typical form expected for a first-order autoregressive process. That is, a decay of auto-correlations to zero, the first-order autocorrelation is around $0.9$ in Figure~\ref{fig:acf_pacf_dp}. The partial autocorrelations become insignificant for lag orders $\geq2$. Hence, the ACF and PACF support the assumption of a stationary first-order autoregressive process for both price-dividend time series. The results become more complicated when we look at the output from augmented Dickey-Fuller and KPSS tests. For the augmented Dickey-Fuller test we do not reject the null hypothesis of a unit root at usual significance levels. For the KPSS the results are mixed. For example, when using the bandwidth selection proposed in \citet{Andrews1991} and the Bartlett kernel we do not reject the null hypothesis that the data is integrated of order zero at a 5\% significance level, while with the \citet{NW1994} selection rule and the Bartlett kernel the null-hypothesis is rejected when applying a significance level of 5\%. 

We claim that these mixed results are at least partially caused by the relatively short time-series dimension. 
\cite{Golez_2018} analyze annual US data from 1629 until 2015.  The autoregressive coefficient for the entire period equals 0.78. However, it becomes more persistent in recent times. For the sample from 1945 until 2015, for instance, the estimate of the autoregressive coefficient is approximately equal to $0.9$. For this larger data set, the authors rejected the null hypothesis of a unit root for the dividend-price ratio at a 1\% confidence level using an augmented Dickey-Fuller test. 
By using this result and the structure of the ACF and PACF we follow the literature \citep[see, e.g.,][]{Cochrane_2008} 
and consider the log {\it Log dividend-price ratio $x_t$}  ratio to be stationary.

\textsc{\begin{figure}[t!]
		\includegraphics[width = 12cm]{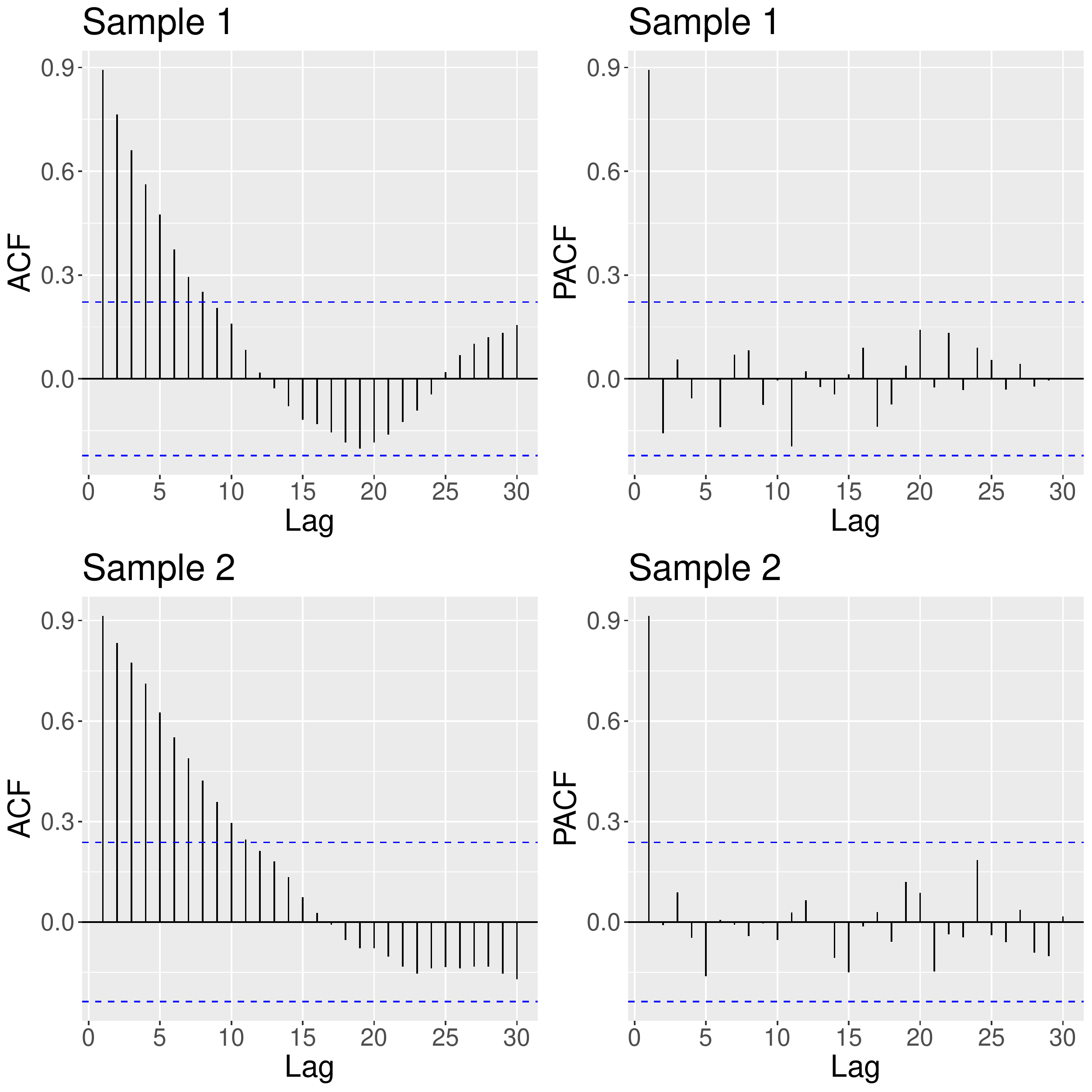}
		\centering
		\caption{ACF and PACF plots for log dividend-price ratio $x_t$.
  The upper panel is the result for {\em Sample 1} and the lower panel is the result for {\em Sample 2} The left panel is the ACF plot and the right panel is the PACF plot. {The dotted lines show 95\% confidence bands.}}
\label{fig:acf_pacf_dp}
\end{figure}}

\subsection[Additional Results]{Additional Results: Estimates of the Marginal Posteriors of $\beta$} \label{app:financial_data_results}

Figure~\ref{fig:sample_1_2_beta} presents the prior and estimated marginal posterior densities  $\widehat{p} \left( {\beta} | \mathbb{D}_{T,\ Sample \ 1} \right)$ for the empirical \textit{Sample 1} (left panel) and
$\widehat{p} \left( {\beta} | \mathbb{D}_{T,\ Sample \ 2} \right)$ for
\textit{Sample 2} (right panel), respectively. The posterior for \textit{Sample 1} has a larger spike at zero than the prior distribution that is in a line with the estimated Bayes Factor being larger than one in Table~\ref{tbl:beta_results}. On the other hand, for \textit{Sample 2} the posterior distribution is slightly smaller than the prior at zero indicating that there is weak evidence of predictability. In addition, the posterior distributions can be compared to the results in Figure~\ref{fig:beta_avg_posterior} for the simulated data, where visually the posterior for \textit{Sample 1} is more representative for \textit{DGP 0}, while the posterior distribution for \textit{Sample 2} is closer to the median of the estimated posterior.

\textsc{\begin{figure}[t!]
		\includegraphics[width = 15cm]{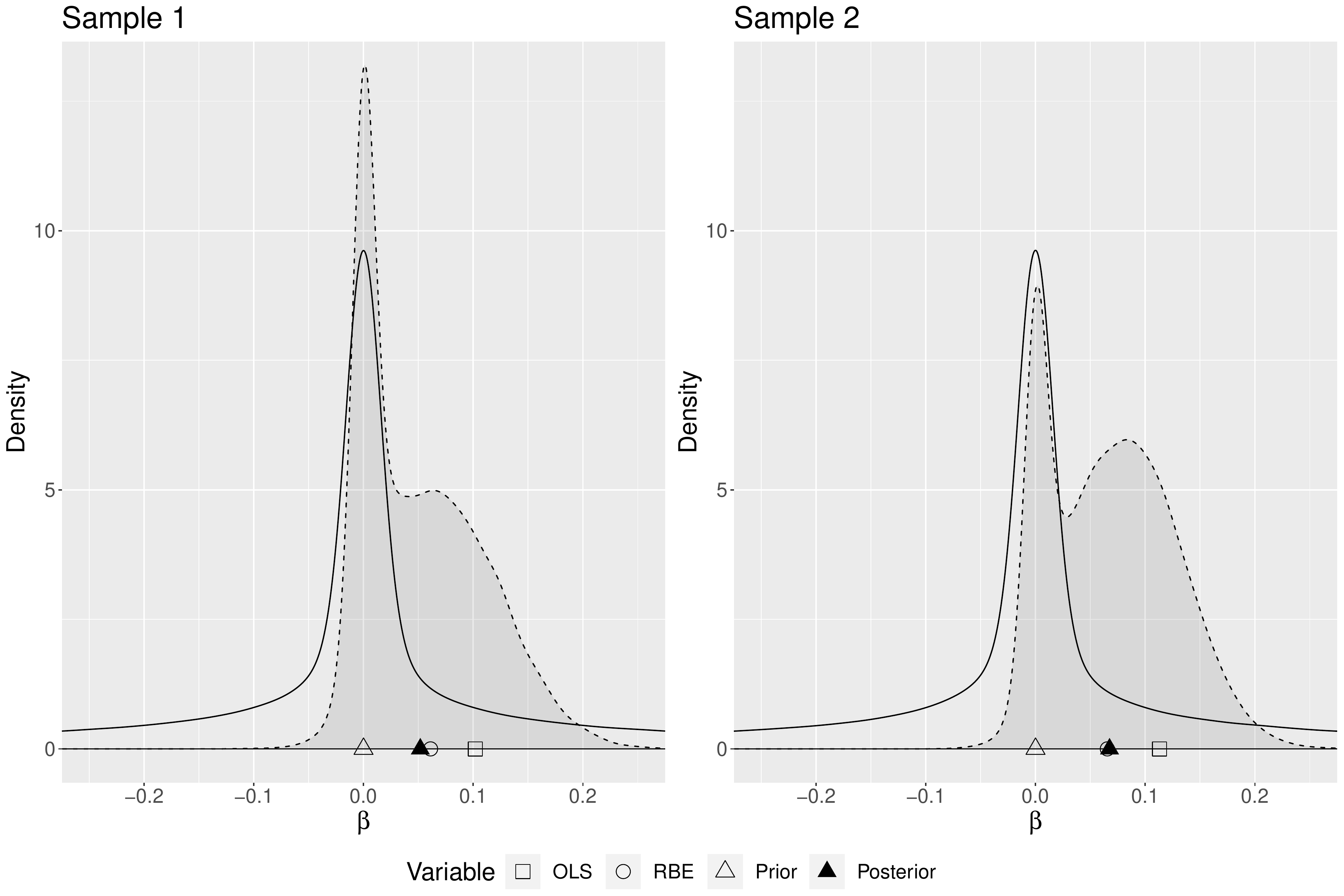}
\centering
\caption{Prior and posterior densities for $\beta$ and point estimates for BAY, OLS and RBE. The solid line is the prior distribution and the dashed line with the filed area stands for the posterior distribution. The empty triangle is the prior mean that equals zero and the filled triangle is the estimated mean of the posterior distribution $\widehat{\beta}^{BAY}$, the empty square is the OLS estimator $\widehat{\beta}^{OLS}$
and the empty circle is the RBE estimator $\widehat{\beta}^{RBE}$. The left panel describes the result for \textit{Sample 1}, whereas the right panel describes the result for \textit{Sample 2}.}
	    \label{fig:sample_1_2_beta}
\end{figure}}

\subsection{Trace Plots and Effective Sample Sizes for the Empirical Data} \label{app:financial_data_traceplots}

In this section, we present the convergence analysis for two samples of the financial data discussed in Section~\ref{sec:financial}. We present the trace plots for $\beta, \phi, \psi$ in Figure~\ref{fig:traceplots.sample} and estimate the effective sample size ${M}_i^{eff}$ in Table~\ref{tbl:ess_1} (where for the parameters $\beta,\psi$ and $\phi$ the corresponding estimates
$\widehat{M}_i^{eff}$, $i \in \{ Sample \ 1, \ \ Sample \ 2  \}$, are abbreviated by ESS $(\beta)$, 
ESS $(\phi)$, and ESS $(\psi)$). 
The details about the convergence analysis are in Section~\ref{app:mcmc_details_convergence}. Visually we conclude that our MCMC sampler  exhibits  good mixing properties for all parameters which is supported by the ESS results.

\textsc{\begin{figure}[t!]
		\includegraphics[width = 12cm]{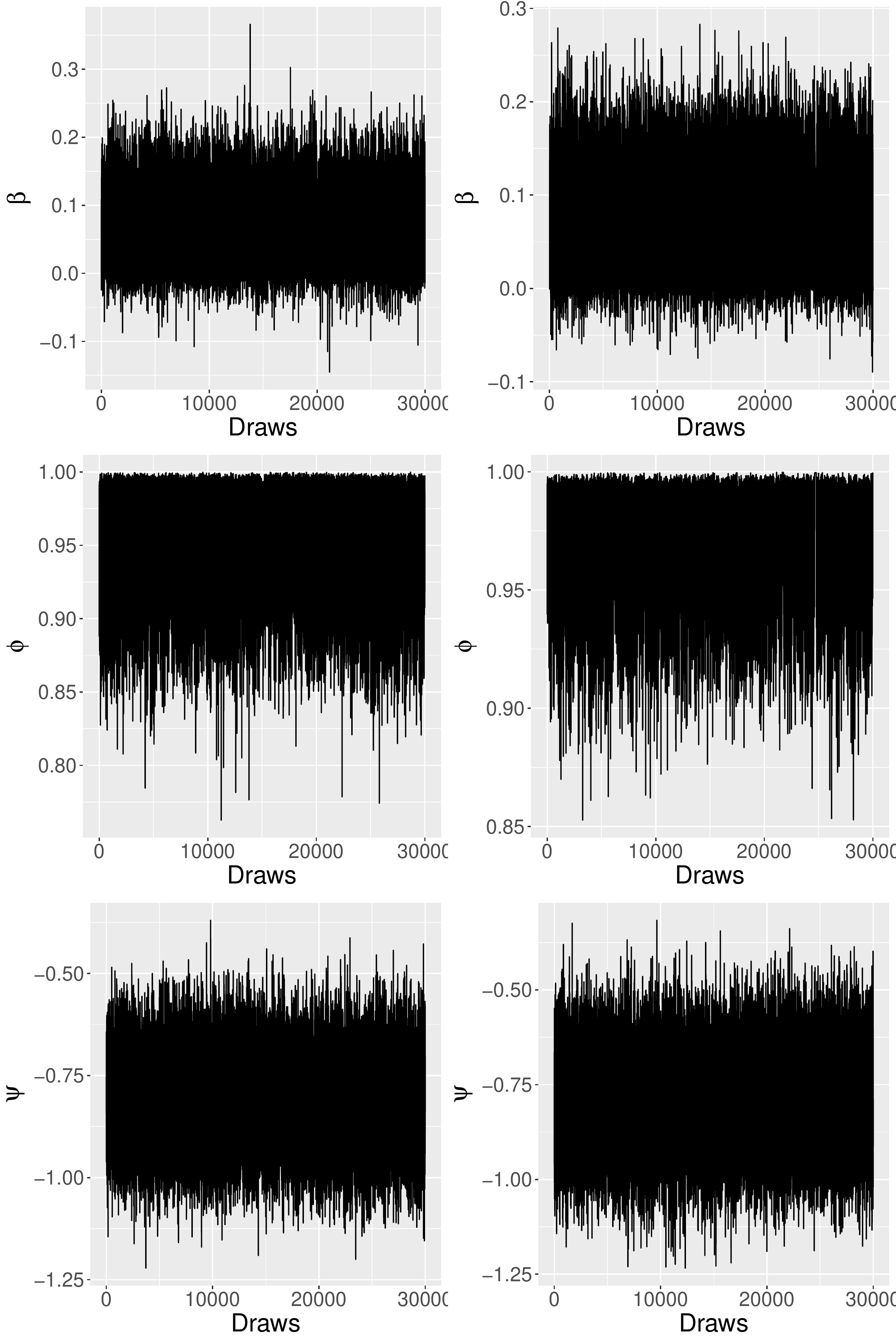}
		\centering
		\caption{Traceplots for the parameters $\beta, \phi, \psi$. The left panel is for \textit{Sample 1} and the right panel is for \textit{Sample 2}}
		\label{fig:traceplots.sample}
\end{figure}}

\begin{table}[!htbp] 
\centering 
\caption{Effective Sample Size for the parameters: $\beta, \phi, \psi$ for the different samples} 
	\label{tbl:ess_1} 
\begin{tabular}{lcccc} 
\toprule
Sample & ESS $\beta$ & ESS $\phi$  & ESS $\psi$  & Length \\ \midrule
\textit{Sample 1} & 11\,868 & 13\,329  & 30\,001  & 30\,001 \\ 
\textit{Sample 2} & 8\,552 & 5\,927  & 28\,745 & 30\,001 \\ 
\bottomrule
\end{tabular} 
\end{table}

\end{document}